# Intelligent Manufacturing Support: Specialized LLMs for Composite Material Processing and Equipment Operation

Gunnika Kapoor[1], Komal Chawla[1], Tirthankar Ghosal[1], Kris Villez[1], Dan Coughlin[1], Tyden Rucker[1], Vincent Paquit[1], Soydan Ozcan[1], Seokpum Kim[1]

[1] Oak Ridge National Laboratory, Oak Ridge, TN, USA

**Abstract**

Engineering educational curriculum and standards cover many material and manufacturing options. However, engineers and designers are often unfamiliar with certain composite materials or manufacturing techniques. Large language models (LLMs) could potentially bridge the gap. Their capacity to store and retrieve data from large databases provides them with a breadth of knowledge across disciplines. However, their generalized knowledge base can lack targeted, industry-specific knowledge. To this end, we present two LLM-based applications based on the GPT-4 architecture: (1) The Composites Guide: a system that provides expert knowledge on composites material and connects users with research and industry professionals who can provide additional support and (2) The Equipment Assistant: a system that provides guidance for manufacturing tool operation and material characterization. By combining the knowledge of general AI models with industry-specific knowledge, both applications are intended to provide more meaningful information for engineers. In this paper, we discuss the development of the applications and evaluate it through a benchmark and two informal user studies. The benchmark analysis uses the Rouge and Bertscore metrics to evaluate our models' performance against GPT-4o. The results show that GPT-4o and the proposed models perform similarly or better on the ROUGE and BERTScore metrics. The two user studies supplement this quantitative evaluation by asking experts to provide qualitative and open-ended feedback about our model's performance on a set of domain-specific questions. The results of both studies highlight a potential for more detailed and specific responses with the Composites Guide and the Equipment Assistant.

Keywords: AI, Large Language Model, Composites Manufacturing, Retrieval-Augmented Generation, Workforce Development, Learning Manufacturing Knowledge

## 1. Introduction

With each year, LLMs increase in size, and thus become more sample efficient [1], allowing for new and better capabilities. The parameter sizes of several widely-used models are: ELMO (94 million parameters), BERT-Large (340 million parameters) [2], Megatron-LM (8.3 billion parameters) [2], Turing-NLG (17.2billion parameters) [2], GPT-3 (175 billion parameters) [2], Megatron-Turing NLG (530 billion parameters) [2], Mistral (7.3 billion parameters) [3], and Llama 3.1 (405 billion parameters) [4]. Researchers have adapted these models for specialized purposes in artistic and scientific fields [5–10]. Interacting with them often feels similar to interacting with humans, given their humanlike style of conversation and potential to critique and refine their thought processes [11]. Composites material is vast, but often disorganized. So, should a user seek to find information on a composites material or manufacturing process without the use of a tool such as an LLM, it might be difficult to find a relevant source or to find the right approach to take, among many, to solve a problem. Additionally, experts have practical experience that gives them a unique perspective with which to identify an effective solution to a problem. Consulting with an expert along with reliable and relevant information would provide the potential for optimal support.

Additionally, when working on the manufacturing floor with specialized machines, technicians and researchers often have to refer back to machine manuals. The process may be hindered due to difficulty in searching through the extensive material to find a specific use case or machine specifications. Using an LLM, this information may be found efficiently.

To these ends, our work makes the following contributions:

• Domain-Specific LLM Applications: We introduce two LLM-based applications that combine the ability of a model to parse and process large datasets with targeted, industry-specific knowledge about composites. The Composites Guide expands on the generalized database of GPT-4 with an additional database of reputed composites articles and magazines and provides users with contact information for experts and businesses in the field whom users can contact for further guidance. The Equipment Assistant utilizes the advanced capabilities of LLMs to utilize specific source material in order to guide users in the operation of manufacturing tools and on material characterization.

• Benchmark Evaluation: We compare the two LLM applications, the Composites Guide and the Equipment Assistant expanded versions of GPT-4, to GPT-4o in order to have objective comparisons between

This manuscript has been authored by UT-Battelle, LLC under Contract No. DE-AC05-00OR22725 with the U.S. Department of Energy. The United States Government retains and the publisher, by accepting the article for publication, acknowledges that the United States Government retains a non-exclusive, paidup, irrevocable, worldwide license to publish or reproduce the published form of this manuscript, or allow others to do so, for United States Government purposes. The Department of Energy will provide public access to these results of federally sponsored research in accordance with the DOE Public Access Plan (http://energy.gov/downloads/doe-public-access-plan)

the capabilities of GPT-4o and the proposed models. The metrics used in order to conduct these comparisons are Chin-Yew Lin's ROUGE metric [12] as well as the BERTScore discussed in Zhang et al [13] Both metrics are automatic evaluation metrics for text generation. The metrics compare text generations to ground truths in order to determine the quality of a generation. The ground truth that was used for the purposes of evaluating the Composites guide was created using a human expert validated dataset of QA pairs generated with NotebookLM based on materials sciences papers. The ground truth QA dataset for the purpose of evaluating the Equipment Assistant was created with a NotebookLM based on targeted machine manuals.

- User Study Evaluation: We show the results of two user studies that evaluate the performance of our applications. The first benchmarks the Composites Guide's performance against that of GPT-4o and analyzes whether our application's curated selection of composites literature and supplementary resources such as contact information for experts and businesses who work in the field improves the level of support users experience with composites-related questions when compared to GPT-4o. The second collects likert-scale-based and open-ended feedback on a selection of prompts in order to evaluate the Equipment Assistant's responses.

In the following sections, we discuss previous work conducted in LLM research and introduce concepts that are intrinsic to our applications' development. We then show the results of the benchmark of our models against Notebook LM and of the two user studies displaying their performance against GPT-4o. Finally, we end with a discussion of potential future applications of both of our applications and their limitations.

**1.1. Previous Work.**

The present selection of LLMs both commercially available and in development spans a wide range of disciplines. For example, Chakrabarty et al. and Zhou et al. seek to aid users with various portions of the creative writing process [5, 6]. Zhou et al. encourage users to rewrite AI-generated text and perform deep thinking [6]. Alternatively, Ning et al. experiment with audio. MIMOSA is a novel human-AI collaborative tool that generates spatial audio effects from videos with monaural audio [7]. Additionally, Bryan-Kinns et al. investigate the use of an AI plugin trained in one genre of music by musicians specializing in a different genre of music [8]. Still others work to create customized personal assistants. Ye et al., for instance, create AI-Yo: a fashion chatbot that curates a selection of fashion styles for users based on the principles of imitation theory, fashion psychology, and recent fashion trends [9].

Some of these developments extend to the composites sphere. Recent projects have explored accurately predicting atomization energies [10], creating concrete formulations that are less CO2-intensive [10], and performing simulations [14] or planning processes [15] to solve problems [15] and predict how composites will perform under set criteria [14]. One project detailed by Jablonka et al. takes this a step further, using a large database to determine the proximity between a vector representation of a set of molecules with a vector representation of a hydrogen carrier molecule [10]. Finally, Gupta et al. pretrain SciBERT to establish a data repository that can be mined for materials science sources. [16].

Industry also has released its own solutions for AI-based assistance in manufacturing. Lee and Su [17] introduced a comprehensive framework for industrial large knowledge models (ILKMs) in the context of Industry 4.0 and smart manufacturing. They explored how ILKMs can integrate diverse knowledge sources to enhance decision-making, optimize processes, and facilitate intelligent automation in manufacturing systems. The framework was designed to enhance the interoperability and adaptability of smart manufacturing systems by harnessing data and knowledge across multiple domains. Furthermore, companies like Dassault and Siemens have highlighted the value of AI in the manufacturing sector through the implementation of AI-driven solutions. These AI-powered platforms integrate with digital twin technology, predictive maintenance systems, and advanced analytics to optimize production workflows, minimize downtime, and enhance quality control. Dassault's AI tools, embedded in its 3DEXPERIENCE platform, support smart design and simulation, while Siemens' AI solutions, within the Siemens Xcelerator ecosystem, utilize machine learning and automation to streamline complex manufacturing processes. As AI technology continues to evolve, these intelligent assistants are becoming crucial for modern smart factories, enabling data-driven decision-making and driving innovation within the industry. This work aims to complement the ongoing digital transformation in manufacturing, assisting technicians in becoming proficient with modern technologies more effectively.

**1.2 Domain Specific Model Training**

To boost LLM performance on domain-specific queries, several strategies are available, with Retrieval Augmented Generation (RAG) being one of the most prominent. RAG combines the strengths of retrieval-based methods and generative models, allowing the model to generate responses based on both its pre-existing knowledge and real-time, relevant external information [18]. During RAG, the documents initially go through an indexing phase where they are processed to create embeddings to be stored in a vector database for later access [19]. After a user enters a query, the model uses the query to retrieve relevant information from this external knowledge source [18] by calculating the similarity between embeddings of the question and embeddings of the document [19]. The information returned from the retrieval process is then further processed to highlight the most important results and remove irrelevant content [19]. The output of the retrieval process is added to the original prompt and then fed into the LLM to generate the final response [18].

Fine-tuning is another common strategy for domain-specific training. It helps guide the model to generate responses that align with a desired or expected response [20] and to improve its accuracy in a particular field [21]. During fine-tuning, a model either undergoes supervised learning or unsupervised learning, depending on whether or not the data the model is trained on is labeled or unlabeled, respectively [22]. Prompt engineering also plays a key role, helping to craft inputs that guide the model to produce more relevant outputs based on user needs [23]. Knowledge injection is yet another approach, where external knowledge bases, such as knowledge graphs or structured data, are integrated into the model to improve domain expertise [24]. Transfer learning can be used to adapt pre-trained models to specific tasks with fewer resources, leveraging knowledge from a broader context [25]. Ensemble methods, where multiple models or techniques are combined, can improve robustness and performance on specific queries. In addition, domain adaptation techniques can be used to tune the model for specific linguistic or terminological nuances found in each domain. Together, these strategies—RAG, fine-tuning, prompt engineering, knowledge injection, and others—offer a comprehensive approach to enhancing LLM performance in specialized domains.

Use in GPT: Both the Composites Guide and Equipment Assistant use a form of RAG to access information contained in curated composites databases to supplement their general knowledge. When a user enters a query, either semantic search or document review is used to identify relevant information. In semantic search, only relevant embeddings are returned with the prompt, while in document review, entire short documents or portions of larger documents are returned with the prompt [26].

For our models, RAG was preferable to fine-tuning because it allowed us to supplement our model's knowledge base with industry and application-specific information, improving the model's accuracy and relevance. Unlike traditional LLMs, which rely solely on pre-existing training data, RAG retrieves relevant documents to generate informed responses, keeping the model up-to-date in fast-evolving fields. It is especially effective for specialized or long-tail queries, where fine-tuning may be limited. Additionally, RAG dynamically retrieves knowledge, handling niche topics more effectively while reducing the need for extensive fine-tuning, saving time and resources. Its scalability and flexibility in retrieval techniques make it adaptable to various domain-specific queries, enhancing accuracy and handling ambiguity [18].

## 2. Design of the Applications

Both of our LLM applications are built on OpenAI's GPT-4 architecture. They combine the standard OpenAI architecture with specific use cases from a corpus of peer-reviewed data to give the model expertise in composites.

Both of the models are hosted on OpenAI's platform and utilize retrieval-augmented generation (RAG) to access an additional database on composites material.

### 2.1 Knowledge Databases

The Composites Guide's dataset consists of a variety of data in the form of txt files, the approximate size of this addition is 2 MB, including:

- Articles published by researchers in the composites field
- Magazines sourced from reputed composites journals
- Contact information for researchers in the composites field
- Businesses that operate in the composites industry
- A summary of a collaboration program known as the ORNL Tech Collaboration Program

The Equipment Guide's dataset consists of articles, machine manuals for thermoformer and injection molder, and literature related to composites manufacturing in the form of txt files, pdf files, and docx files, the appoximate size of this addition is 18 MB. Through augmenting the original database of GPT 4o with composites databases, both models are able to construct more precise outputs tailored to users' specific needs. A portion of the training dataset used for the Composites Guide is shown in Fig. 1.

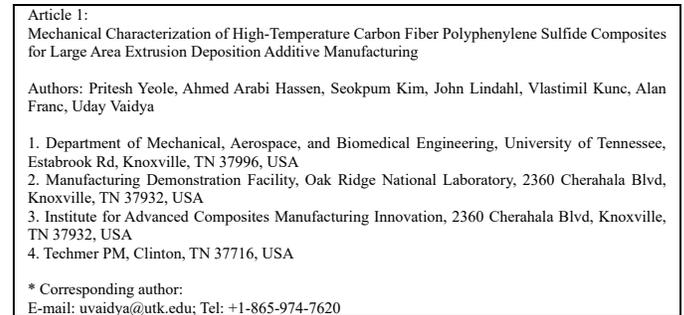

*Fig. 1 Sample Passage from Composites Guide Database. A brief excerpt from one of the txt files that the Composites Guide contains in its database.*

### 2.2 Model Architecture

OpenAI's GPT models have many capabilities that were essential for this project, such as the ability to parse files and access large databases of knowledge. Excitingly, their GPT-4 [27], GPT-4o, GPT-4o mini, and GPT-4 Turbo models [28] have image processing capabilities [27, 28]. This is an important feature, as it gives users the option of attaching reference images, schematics, or other supporting material with their prompts to help clarify their thought process and provide additional context for their query. For this reason, the Composites Guide and Equipment Assistant are built on the GPT-4 architecture.

Model Features: The models (Appendix C: Fig. 16) support the following capabilities.

- Submitting a text-based query (Fig. 2)
- Uploading an image (Appendix C: Fig. 17)
- Uploading a file (Appendix C: Fig. 18)

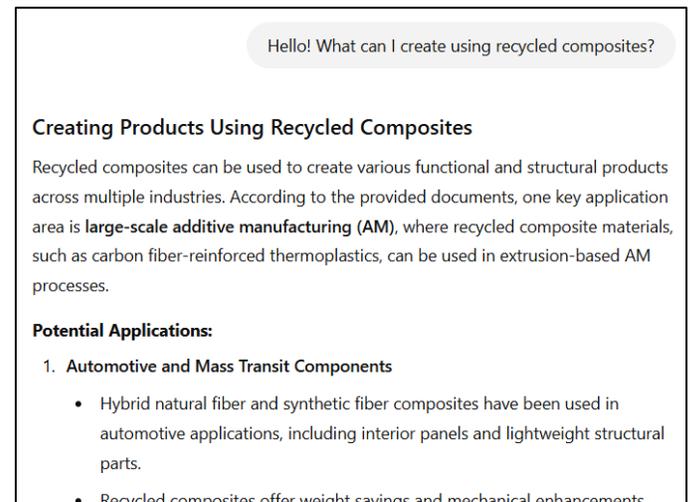

*Fig. 2 An initial portion of the Composite Guide's response.*

## 3. Evaluation

In order to evaluate the performance of the Composites Guide and the Equipment Assistant, we performed both quantitative analysis through a benchmark and qualitative analysis through two user studies. We utilized classical NLP metrics to quantifiably evaluate the proposed models against the baseline GPT-4o model. We validated our results using human evaluation practices. We supplemented these results with insights from two limited, informal user studies. In this study, we asked a set of participants with relevant experience and knowledge to compare our models' performance against that of GPT-4o. This allowed us to get additional detailed feedback regarding their experience with our models.

## 4. Benchmark

### 4.1 Evaluation Dataset

The evaluation dataset, consists of QA pairs that have been generated by NotebookLM. NotebookLM has been chosen as the model to provide this ground truth as it is designed to use only the sources we provide to it. For the generation of the Composites Guide QA pairs, NotebookLM was given a set of 10 composites papers and made 100 QA pairs, after these QA pairs were generated, a human processed the dataset removing generic and nonspecific responses having the final composites QA dataset totaling 63 questions.

The evaluation dataset for the equipment assistant was also generated with NotebookLM. Similarly, for the generation of the Equipment Assistant QA pairs, NotebookLM was given the manuals of the injection molding machine and the MAAC Thermoformer, in order to generate 100 questions for each. The resulting dataset consisted of 200 QA pairs generated by NotebookLM. Samples of both datasets can be found in their respective sections in the appendix G and H.

The metrics, conducted through human evaluation and automated evaluation metrics allow us to adopt holistic approach to evaluating the two proposed models. While the automated evaluation metrics allow us to quantifiably evaluate similarities between the ground truth, the baseline and the proposed model's generation, the human evaluation metrics are valuable for quantitatively evaluating the proposed models against GPT-4o. While similarity is important, this work also highlights the importance of user experience and addressing the needs of users through the modifications that were made.

### 4.2 Method – Automated Approach

The metrics used for the evaluation of the Composites Guide and the Equipment assistant to a non-modified GPT-4o are the ROUGE metric and the BERTScore metric. Both metric suites take in a list of text generations and compare them with the list of ground truths. The average of these scores is taken in order to take a system level evaluation.

The BERTScore and ROUGE metric were chosen in order to evaluate the similarities between the ground truth and the generated responses. While the ROUGE-N suite focuses on similarities between n-grams, sequences of n-adjacent symbols, and the ROUGE-L metric evaluates longest common subsequence similarity, the BERTScore differs by computing similarity using contextual embeddings. Both metrics are useful in evaluating the overall similarity between test cases and the ground truth.

### 4.3 Results from Automated Approach

The metrics used in evaluating the proposed models compared to the baseline GPT 4o are the Rouge metric and the Bertscore. The metrics test the similarities between the generation of the test models and the gold standard. The gold standard in this study has been created by utilizing NotebookLM to generate source-based question answer pairs centered around composites sciences papers. After this dataset was generated by feeding a sample set of composites research papers to NotebookLM, the QA pairs were then validated by the authors of the composites research papers in order to create a human-backed gold standard. The goal of this gold standard is to support the notion that the proposed model's access to specific user manuals quantifiably benefits LLM generations.

There is not a significant improvement based on the ROUGE metrics for the Composite Guide compared to GPT-4o, figures 19, 20, and 21 depict inconclusive results that do not suggest an improvement on the ground truth dataset. Similarly, the ROUGE metrics of the Equipment Assistant compared to GPT-4o are also inconclusive as seen in figures 22, 23, and 24. Both the Composites Guide and the Equipment Assistant performed similarly to GPT-4o based on the BERTScore Metric as seen in figures 25 and 26.

The main reason for the inconclusive results is that the two metrics, ROUGE and BERTScore, do not evaluate the meaning of the sentences, but instead mechanically find the similarities. Below are two examples.

| Q#30. What is the most important factor in thermoforming besides sheet quality? | |
|---|---|
| A#30. Assuming the thermoformer has good sheet to work with, the next most important factor is its proper heating. | |
| GPT-4o | Equipment Assistant |
| **Proper heating control** is the most critical factor as it ensures uniform material softening and prevents defects. | Besides sheet quality, the most important factor in thermoforming is **consistent and accurate temperature control**.<br>• **Even heating ensures uniform material distribution**.<br>• **Temperature control prevents defects** like thinning, warping, or improper forming.<br>• **Optimized heating zones reduce cycle time and improve part consistency**.<br>Without precise temperature control, **variability in forming results increases**, leading to **higher scrap rates**. |
| Q#31. What is profile heating and what benefits does it have? | |
| A#31. Computer controlled multi-zoned ovens are now capable of profile heating which has the ability to produce a consistent sheet temperature from the perimeter to the center, reduces sag, uses less energy and decreases the heating and cooling cycle times. | |
| GPT-4o | Equipment Assistant |
| **Profile heating adjusts heat distribution across different zones of the sheet,** ensuring consistent material thickness and improved forming accuracy. | **Profile heating** is a method where **different zones of the plastic sheet are heated at different intensities** to optimize material flow.<br>**Benefits include:**<br>• **Better thickness distribution**—prevents **thin spots** in deep-draw areas.<br>• **Reduced stress on material**—improves **part durability**.<br>• **More efficient energy use**—heats only where necessary.<br>• **Minimizes warping and sagging**—ensures **high-quality finished parts**.<br>Profile heating is **especially useful for forming complex geometries**. |

Both answered correctly, with the answers from Equipment Assistant being more detailed and informative. However, with mechanical comparison word by word, they are not similar to the ground truth answers, and both answers were scored poorly for the two questions[1].

## 5. Experts (Human) Evaluation Benchmark

The same QA datasets (100 for each application) that were used in the automated evaluation metrics were given to human experts, and the results are shown below.

*Table 1: Experts (Human) Evaluation Benchmark*

| Q&A for composite materials | |
|---|---|
| GPT-4o | Composites Guide |
| Avg. Score 3.23 | 4.0 |

| Q&A for equipment operation | |
|---|---|
| GPT-4o | Equipment Assistant |
| Avg. Score 4.14 | 4.58 |

\* Rating ranges from 1 (completely incorrect) to 5 (completely correct)

Below is an example QA for which both GPT-4o's answer and Equipment Assistant's answer were rated poorly.

| Q#56. What is the MCF OVERRIDE button used for? | |
|---|---|
| GPT-4o | Equipment Assistant |
| It allows manual control of platen motion to override the default motion control functions. | ● Temporarily bypasses mold close force limits. <br> ● Allows higher force application for certain materials. <br> ● Used only in specific situations where standard settings prevent proper forming. |
| Expert's evaluation | Expert's evaluation |
| Rating 2: Incorrectly explains as manual platen motion control. The answer should be it's for moving the car away from form station during setup | Rating 2: Incorrectly explains as a mold close force override. The answer should be it's for moving the car away from form station during setup |

Below is an example QA for which GPT-4o answered poorly and Equipment Assistant answered correctly.

| Q#26. What precaution should be taken when handling clamp frames? | |
|---|---|
| GPT-4o | Equipment Assistant |
| Ensure that clamp frames are securely locked in position before adjusting them, as sudden movements can cause injuries. | ● Always use two hands when handling clamp frames to prevent dropping. <br> ● Ensure clamps are properly locked before forming begins. <br> ● Watch for pinch points—clamp frames can cause severe hand injuries. |
| Expert's evaluation | Expert's evaluation |
| Rating 3: It mentions securing clamp frames but misses the pinch points and two-handed handling. | Rating 5: Correct answer. |

Below are the overall evaluation comments.

| | GPT-4o | Equipment Assistant |
|---|---|---|
| Strength | • Strong understanding of the manufacturing process <br> • Clear and concise explanations | • More accurate on technical specifications <br> • Better understanding of machine interface and controls <br> • Good explanation of the manufacturing processes <br> • Less critical errors on safety-related questions |
| Weakness | • More frequent factual errors on specific measurements and settings <br> • Several critical errors on safety-related settings <br> • More misinterpretations of UI and controls | • Occasionally provides generalized ranges instead of specific values <br> • Occasionally misinterpret specialized functions |

## 6. Study 1: Composites Guide

### 6.1 User Study Recruitment, Selection, and Design

To evaluate the Composites Guide, we compared its performance with that of GPT-4o through a limited user study conducted among 5 researchers who work at the Manufacturing Demonstration Facility (MDF) at Oak Ridge National Laboratory (ORNL) and have experience working with composites. No demographics information was collected. We generated a total of 6 prompts to evaluate the model. They can be organized into categories as follows:

- Questions a beginner to composites might ask: Prompt 1
- Questions for targeted/specific applications: Prompts 2 - 5
- Questions an expert in composites might ask: Prompt 6

For each prompt, 2 responses were generated. One response was generated by the Composites Guide and the other was generated by the standard GPT-4o model. The prompts and corresponding responses are shown in Appendix A (Fig. 4 - Fig. 9). In order to prevent bias, users were not made aware of the category names and all responses were labeled generically as "Response 1" and "Response 2". Additionally, the models that generated the responses were alternated with each prompt. After selecting the response they preferred for each prompt, users were invited to answer an optional question and explain the reasoning behind their selection.

### 6.2 Statistical Analysis

The results from Fig. 3 indicate which model's response users preferred for each prompt. Users preferred GPT-4o's responses for Prompts 1, 4, and 5, and preferred the Composites Guide's responses for Prompts 2, 3, and 6. The prompts with the greatest margin of support for one model were Prompts 5 and 6, for which GPT-4o and the Composites Guide were found to provide better responses, respectively. When analyzing the user's written responses in conjunction with their choices for each prompt, a couple of interesting themes are identified, which are discussed in the following section, User Insights.

---

[1] In contrast to the poor results from the automated metrics, a human expert we selected for thermoforming process rated 5 for both GPT-4o answers and Equipment Assistant answers for both Q#30 and Q#31.

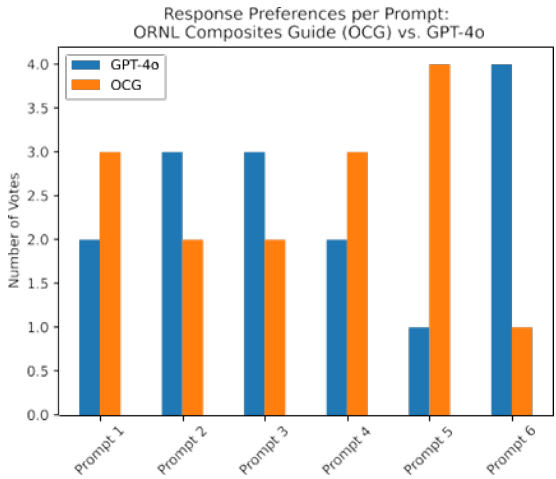

*Fig. 3 User Study Results. The number of votes a model's generated response received for each prompt.*

## 6.3 User Insights

### 6.3.1 GPT-4o gives a general overview, but the Composites Guide provides the details.

The question of which model provides better support for beginners to composites is a subjective one, since it depends on the user in question and on their learning style. For example, GPT-4o performed better for Prompt 1 than the Composites Guide, because most users found that it was *(User 2) more general in nature for the beginner,* and had *(User 1) more concept and easier transition from very basic...to more detailed info.*

However, Users 4 and 5 disagreed. *(User 4) response 1 [GPT-4o] sounds too general, and too much like a road map.* This sentiment was shared by User 5. *(User 5) The first response [GPT-4o] is more generic. For starters, the information provided and resources mentioned in response 1 would be more useful.*

This indicates that whether or not a general response is preferred depends on the user. However, at least for this prompt, GPT-4o's response can be classified as general and, according to the majority of users, favorable.

Those who preferred the Composites Guide appreciated its inclusion of information such as definitions and terms. *(User 3) From a personal standpoint, if I'm asking a question like this a great starting point is understanding the terminology.*

We can generalize and say that GPT-4o appears to provide responses that would better explain the concept if users were hoping to get a general overview of the topic. However, if they have some familiarity with the subject area or prefer more detailed responses, then the definitions and specific information found in the Composites Guide's answers could be helpful.

This is confirmed by User 2's response to Prompt 5. *(User 2) I would prefer response 1 [GPT-4o] at first, then if I wanted to drill down to composite solutions response 2 [Composites Guide] would be more appropriate.*

### 6.3.2 Preference for model depends on dataset, timing, user

For the remainder of the responses, users' preferences came down to the format of the response. Some preferred the structure of GPT-4o's responses, but others identified key pieces of information that were in the wrong section or missing in its responses. When discussing Prompt 3, both User 2 and User 3 preferred the Composites Guide. Interestingly, User 3 mentions, *(User 3) I like that the second response [Composites Guide] gives you more than one option where response 1 [GPT-4o] only makes it seem like there are fewer ways to fabricate the piece.* User 1 adds for Prompt 2 that *(User 1) The question is actually answered ("Yes, ...") in response 1 [Composites Guide]. There is no clear answer in Response 2 [GPT-4o].*

On the other hand, User 4 says *(User 4) I didn't like neither response actually. They do not answer the question. They are giving way more information then asked. But 2nd [GPT-4o] one sounded more direct relatively, that is [why] ... I selected it.*

All of these responses suggest that preference for model is not only dependent on the content of the response, but also on the format. This is clearly seen in User 3's response. The language the Composites Guide uses encourages users to explore multiple options, and likely to either ask for clarification or to look at the additional contact information and resources the model provides.

However, the format of a response often varies from prompt to prompt. At times, the response may also be inaccurate, or as was the case for User 4, unfavorable. One reason for this variance could be the size of the dataset. GPT-4o has a significantly larger number of parameters. This gives it knowledge about more topics than the Composites Guide. So, if our model had less expertise on a subject, it was likely to provide less support. This led to interesting results. User 5 found that the *(User 5) additional resources and list of items in response 1 [GPT-4o] makes it easier for adoption.*

The Composites Guide is specifically instructed to provide these resources, while GPT-4o might not be. However, even with these instructions, it did not include it in its response, while GPT-4o did. So, the choice of which model is the better one might also depend on the size of the dataset and when the response was generated.

## 7. Study 2: The Equipment Assistant

## 7.1 User Study Recruitment, Selection, and Design

In order to evaluate the Equipment Assistant, we conducted an additional informal user study consisting of 3 participants who work at the Oak Ridge National Laboratory (ORNL) under the Manufacturing Demonstration Facility (MDF) with relevant experience working with manufacturing equipment. No demographics information was collected. As all of the equipment manuals uploaded to the model are ORNL-specific equipment manuals, rather than evaluating it against GPT-4o, which may not have such specialized information, we opted to get more general feedback about the responses from our target end-users, equipment operators. We generated a total of 6 prompts. They can be organized as follows:

- How to Use: Prompts 1 and 6
- Troubleshooting: Prompts 2 and 5

- Safety Guidelines and Machine Specifications: Prompts 3 and 4

The prompts and corresponding responses are shown in Appendix B. For each prompt, a response was generated by the Equipment Assistant. Users were first asked to rate each response on a 5-point Likert Scale. After that, they were asked to answer another question to provide further feedback and suggestions.

## 7.2 Statistical Analysis

Table. 2 depicts results of the informal user study. Overall, Equipment Assistant outputs received high scores of 4 and 4.66. Prompts 3 - 6 received the highest scores. Themes derived from users' written responses will be discussed below.

*Table 2 Average ratings provided for each prompt generated during Study 2.*

| Prompt Number | Average Rating |
| --- | --- |
| Prompt 1 | 4 |
| Prompt 2 | 4 |
| Prompt 3 | 4.66 |
| Prompt 4 | 4.66 |
| Prompt 5 | 4.66 |
| Prompt 6 | 4.66 |

## 7.3 User Insights

*7.3.1 Users prefer concise and detailed responses.*

Overall, users participating in the study thus evaluated the responses in terms of validity and conciseness. While some users were satisfied with the generations: *(User 2) The explanation looks good to me.* Others suggested areas where more detail could improve the responses. For example, for Prompt 2, User 1 mentioned *(User 1) to improve this response, I would prefer to also see the electrical specifications that would be required to know if a replacement component is needed.*

Additionally, for Prompt 1, they mentioned that *(User 1) additional information regarding the rationale to determine an appropriate clamp force would be ideal.* One the other hand, some users sometimes found that the responses provided extra, unnecessary detail. For Prompt 2, User 3 said *(User 3) "it is essential to ensure proper surface contact with the nozzle body for accurate heat transfer and effective operation." This part of the feedback seems irrelevant* These responses indicate that users prefer varying degrees of detail and focus on different aspects of the application's generations when considering what makes a response a useful one.

*7.3.2 The Equipment Assistant provides a good starting point.*

The responses suggest that while the application offers a good point of reference, users appreciate the ability to refer back to the source material to verify responses. This can be seen in User 2's response to Prompt 6. *(User 2) Response gives a good starting point for the sheet size.* with User 1 adding in response to Prompt 5, *(User 1) I really like the thorough response with further references.*

These responses suggest that an important aspect of the responses is that it directs the user to another reference, encouraging them to validate the information in the response and perform further research if desired before taking actions.

## 8. Discussion

### 8.1 Potential Application

As the Composites Guide and Equipment Assistant have targeted and domain-specific knowledge bases, they have the potential to provide a better user experience as they can be modified for more specific use cases. Below are potential benefits.

**Providing new options:** Since composite materials are not well understood by many engineers and designers, it is intended that these frameworks would provide the guidance needed to explore composites solutions for design and engineering challenges.

**Saving Time:** Training employees on a new concept can be a time-consuming process, due to the time spent gathering learning resources such as relevant papers and articles and time spent providing direct guidance. One way the models can help with this is by providing support for basic questions or general knowledge about the subject. This reduces reliance on the mentor to provide background and introductory information.

**Offloading Work:** Certain teams may have specialized knowledge in a subject area that is essential to the functioning of other groups. This can cause those teams to frequently address recurring questions. By updating the models' knowledge database with general information to answer those questions, they can be directed to the model, instead. This reduces the workload on those teams and allows them to be a resource for more specialized queries.

### 8.2 Future Work

*8.2.1 Composites Guide*

When responding to Prompt 6, User 2 gave the following response: *(User 2) Slight edge to Response 1 [Composites Guide]. However neither response uses the method that I have seen used commercially which is SMC or bulk molding compound which is interesting. Perhaps with more input examples it would have found those options.*

Since the model is intended to contain material written by experts in the composites field, its responses should reflect methods that are known to be successful.

Future work should test the accuracy of the Composites Guide's responses in a real-world situation. Such a study might ask experts to provide a solution that has consistently and correctly solved a composites problem. The solutions can then be compared with the application's suggestions to analyze the application's response for viability.

*8.2.2 Equipment Assistant*

A key insight from Study 2 was that the application provides a good starting point. This provides a critical path for future work that emphasizes the creation of an application that serves as a supplemental material to the source material. Users appreciated the model providing citations and suggesting searching for further details in the source material. This could be applied in the future by supplementing the current architecture with a RAG architecture specifically centered around this use case that emphasizes providing relevant citations and references.

## 8.3 Further Discussions

While modifying the Composites Guide and Equipment Assistant, many aspects of the intersection of artificial intelligence and the material sciences were highlighted. Safety and error mitigation are critical aspects pertaining to the use of AI in the materials sciences. In order to ensure system robustness, mitigate errors, and promote safety the equipment assistant specifically is prompted in order to provide safety information as well as operating instructions. The composites guide helps to ensure safety and error mitigation by providing contact information and citing sources to experts in the field. While these are first steps to helping ensure safety and mitigate errors, further work can be done in order to ensure that LLMs are utilized safely and responsibly by all users.

## 8.4 Limitations

### 8.4.1 File Creation

Creating the txt files that form a large portion of the models' composites database relied to a large extent on manual processing. The original files were in the form of pdf files, and the conversion to txt files stripped away a majority of the formatting and all of the non-tabular figures, resulting in large portions of unreadable text. To resolve these errors, we manually processed the files to correct and remove portions the model would not be able to parse. As such, the process is subject to human error: portions of the files may still contain invalid formatting or other errors that affect the model's ability to parse the files. This can have an effect on the responses it generates, as the corrupted portions may cause a loss of information or be misinterpreted by the model. To generate more consistent results, future processing may instead begin with an automated method of processing that is then supplemented by manual revision and verification of the results.

We explored alternative processing methods such as converting from a pdf file to a markdown file using pandoc, or using Adobe Acrobat to convert from a pdf file to a txt file. While the markdown file retained more of the structure of the pdf file than the txt file, both of the automated versions still resulted in unreadable characters and invalid structure. As such, it would still require manual processing to correct errors and would still result in a loss of information. As such, future methods must still be explored.

### 8.4.2 Evaluation Dataset

The evaluation dataset, consists of QA pairs that have been generated by NotebookLM, followed by experts' verification. NotebookLM has been chosen as the model to provide the ground truth answers as it is designed to use only the uploaded documents. For the generation of the Composites Guide QA pairs, NotebookLM was given a set of 10 composites papers and made 100 QA pairs, after these QA pairs were generated, human experts processed the dataset removing generic and nonspecific responses having the final composites QA dataset totaling 63 questions.

The evaluation dataset for the equipment assistant was also generated with NotebookLM. Similarly, for the generation of the Equipment Assistant QA pairs, NotebookLM was given the manuals of the injection molding machine and the Thermoformer, in order to generate 100 questions for each. The resulting dataset consisted of 200 QA pairs generated by NotebookLM and human experts verified the dataset.

**Composites Guide:** Due to copyright limitation, the current dataset contains institutional bias, with 20 out of 123 articles written by researchers affiliated with Oak Ridge National Laboratory (ORNL). Additionally, most of those who agreed to be included in the system are ORNL researchers. As such, the Composites Guide often suggests those experts or mentions their papers in its responses. This might take place even when papers from other sources or experts with alternative affiliations could provide greater support, due to the saturation of information from that source. We plan to incorporate a more balanced representation of literature and expert contacts.

**Equipment Guide:** The model's accuracy for manufacturing processes is highly dependent on the user manuals. Inquiries that require information not included in the manual may lead to too general answers. Therefore, we plan to add additional documents such as polymer theory books or manufacturing process handbooks to improve the knowledge base.

### 8.4.3 Industry Models for Benchmarking

Although there exists industrial AI assistance tools, due to the time and resource constraints, we were unable to obtain licenses for these industrial AI-assistants such as those made by Dassault and Siemens. A future direction of this work would be to obtain these licenses in order to make comparisons beyond the baseline GPT-4o model. Future work could address this limitation by conducting the same experiments and utilizing the same metrics in order to evaluate the proposed models against assistants used in industry.

### 8.4.4 Benchmark

The benchmarks and human evaluation metrics provide a starting point for further study. The benchmarks are limited by their focus on measuring similarity and therefore some semantic similarity measures are needed. In order to rectify this necessity, we attempted to bridge the gap through human evaluations but human evaluations also carry possible biases and person-specific insights. Further work is intended to address these limitations, including a more extensive metrics suite tailored to semantic similarity and further evaluating how user experience is altered by the proposed model.

### 8.4.5 User Study

**Composites Guide:** Given the limited timeline and scope of the study, the participants chosen were among those who contributed to the model's dataset. Thus, their responses were likely partially biased towards the Composites Guide. There is also uncertainty due to the small sample size. In the future, we hope to validate the results of our study by running another trial with a larger set of participants.

**Equipment Assistant:** As the scope and timeline were limited for this study, the sample size of the participants was small and limited to in-house users of the equipment. In a future work, we intend to confirm the results of our study by conducting trials with an increased number of participants.

*8.4.6 Reliability and Safety Validation*

LLMs can produce plausible-sounding but incorrect information, a.k.a. hallucinations, and therefore, this application is intended to serve as a decision support tool but not an autonomous decision maker. In case of safety-related inquiries, it is important to include human experts to review the recommendations. By providing sources for the responses the models generate and, in the case of the Composites Guide, by providing expert sources for further contact, the models allow the user to verify the provided responses or to conduct further research should they desire to do so.

## 9. Conclusion

While commonly-used LLMs such as GPT-4o offer information on composites, their responses can lack specificity due to the absence of a specialized composites dataset. The Composites Guide and Equipment Assistant combine the beneficial characteristics of such large language models with a tailored database of composites material and industry-specific information and support for users of all experience levels. As User Study 1 showed, each user has individual needs and expectations from a LLM, and this can inform their preferences for model responses. Depending on the size of the dataset the model is trained on and when a response is generated, the responses can vary. The Composites Guide met several of these requirements. Additionally, Study 2 indicated that users appreciate a model providing citations for generated content and references to allow them to confirm model output and perform further research, both of which were found in the Equipment Assistant. Both the Composites Guide and the Equipment Assistant have the potential to improve access to composites-related material and several aspects of composites manufacturing.

## Acknowledgment

This research is sponsored by the U.S. Department of Energy, Office of Energy Efficiency and Renewable Energy, Industrial Technologies Program, under contract DE-AC05-00OR22725 with UT-Battelle, LLC.

## Appendix A: Prompts Used in User Study 1

Prompt 1:

I don't know anything about polymer composite manufacturing, although I have some machining experience. I want to learn about polymer composite manufacturing. Where can I start?

| Response 1 | Response 2 |
|---|---|
| Getting started with polymer composite manufacturing involves understanding the materials, processes, and applications involved. Given your machining experience, you might find some aspects easier to grasp. Here are some steps and resources to help you get started: <br><br> 1. Basic Understanding of Composites <br> • Definition: Composites are materials made from two or more constituent materials with different physical or chemical properties. When combined, they produce a material with characteristics different from the individual components. <br> • Types: Polymer composites typically involve a polymer matrix (such as epoxy, polyester, or vinyl ester) and a reinforcing material (like glass, carbon, or aramid fibers). <br><br> 2. Foundational Knowledge <br> • Books: <br> "Introduction to Composite Materials" by Stephen W. Tsai and Hyer <br> "Composite Materials: Science and Engineering" by Krishan K. Chawla <br> "Principles of Composite Material Mechanics" by Ronald F. Gibson <br><br> • Online Courses: <br> • Coursera and edX offer courses on composite materials and manufacturing. <br> • MIT OpenCourseWare provides free materials for courses like "Mechanical Behavior of Materials" which includes composite materials. <br><br> 3. Understanding Manufacturing Processes <br> • Processes: <br> • Hand Lay-Up: Manual process of laying resin-soaked fiber sheets in a mold. <br> • Filament Winding: Continuous filaments are wound under tension over a rotating mandrel. <br> • Pultrusion: Continuous process where fibers are pulled through a resin bath and then through a heated die to form a continuous composite profile. <br> • Resin Transfer Molding (RTM): Fibers are placed into a mold and resin is injected. <br> • Vacuum Infusion: Resin is drawn into a fiber lay-up using a vacuum. <br> • Autoclave Molding: Uses high pressure and temperature to cure composites, typically in aerospace applications. <br><br> 4. Hands-On Learning and Workshops <br> • Workshops: Many institutions and companies offer hands-on workshops where you can learn the basics of composite manufacturing. <br> • Community Colleges/Universities: Enroll in courses related to composite materials and manufacturing. <br><br> 5. Industry Standards and Best Practices <br> • Organizations: <br> • ACMA (American Composites Manufacturers Association): Offers resources, certifications, and standards. <br> • SAMPE (Society for the Advancement of Material and Process Engineering): Provides conferences, technical papers, and industry standards. <br> • Standards: Familiarize yourself with ASTM and ISO standards related to composites. <br><br> 6. Technical Papers and Journals <br> • Journals: Reading journals like "Composites Science and Technology" or · ournal of Composite Materials" can provide insight into the latest research and developments. <br> • Technical Papers: Access papers through platforms like ResearchGate, Google Scholar, or university libraries. <br><br> 7. Networking and Professional Groups <br> • Forums and Groups: <br> • Join LinkedIn groups related to composite manufacturing. <br> • Participate in forums like CompositesWorld or The Composites Forum to ask questions and share knowledge. <br> • Conferences: Attend conferences and trade shows like JEC World or CAMX to network with professionals and learn about the latest technologies. <br><br> 8. Practical Application <br> • Projects: Start with small projects to apply what you've learned. Consider building simple composite structures or repairing composite items. <br> • Internships/Apprenticeships: Consider working with a company that specializes in composites to gain practical experience. <br><br> By following these steps and utilizing the resources mentioned, you can build a solid foundation in polymer composite manufacturing. Given your machining background, you may find certain processes intuitive, especially those involving tooling and material handling. | To get started in polymer composite manufacturing, given your background in machining, here are some key areas and resources to focus on: <br><br> 1. Understanding the Basics of Polymer Composites <br><br> Polymer composites are materials made from a polymer matrix reinforced with fibers or other fillers to enhance properties. Common types include fiberglass-reinforced plastics (FRPs) and carbon fiber- reinforced polymers (CFRPs). <br><br> 2. Key Manufacturing Techniques <br><br> Learn about the primary manufacturing techniques used in polymer composites: <br> • Extrusion Molding: Involves pushing the polymer composite material through a die to create a specific shape. <br> • Injection Molding: Involves injecting molten composite material into a mold to form parts. <br> • Compression Molding: Involves placing composite material into a heated mold, which is then closed to shape the part. <br> • Additive Manufacturing: Includes 3D printing techniques to build parts layer by layer. <br><br> 3. Industry Applications <br><br> Polymer composites are used in various industries, including aerospace, automotive, construction, and renewable energy. Understanding the applications can help you identify where your machining skills might be particularly valuable. <br><br> 4. Learning Resources and Collaboration Opportunities <br> • Oak Ridge National Laboratory's Tech Collaboration Program (TCP): This program engages industry partners in short-term collaborative projects to accelerate the development of new manufacturing technologies. It provides access to ORNL's experienced staff and unique equipment, facilitating rapid development and adoption of new manufacturing processes. <br> • Industry Suppliers and Manufacturers: Familiarize yourself with companies involved in producing and supplying polymer composites. For example, Creative Composites Group and CMDT Manufacturing are among the many companies specializing in FRP products. <br><br> 5. Expert Contacts <br> Connecting with experts can significantly enhance your learning experience. Here are a few key contacts at ORNL who specialize in polymer composites: <br><br> Dr. Ahmed A. Hassen <br> Group Leader & Senior R&D Staff Scientist, Composites Innovation Group, ORNL Email: hassenaa@ornl.gov <br> Dr. Hassen specializes in high-rate manufacturing of advanced composite structures and strategies for smart manufacturing of polymers and advanced composites. <br><br> Dr. Amber M. Hubbard <br> R&D Associate Staff Member, Sustainable Manufacturing Technologies Group, ORNL Email: hubbardam@ornl.gov <br> Dr. Hubbard focuses on natural fiber-reinforced composite formulation and processing, as well as vitrimer composite characterization. <br><br> 6. Further Reading and Training <br> • ORNL Industry Collaboration Program: Provides resources and opportunities for hands-on experience and collaboration with industry experts. <br> • Professional Organizations: Join organizations such as the Society of Plastics Engineers (SPE) and the Materials Research Society (MRS) to access publications, attend conferences, and network with professionals in the field. <br><br> Starting with these resources and connections will provide you with a solid foundation in polymer composite manufacturing and help you leverage your existing machining experience in this advanced field. |

*Fig. 4 User Study 1 Prompt 1. Response 1 was generated by the Composites Guide Model and Response 2 was generated by the standard GPT-4o model.*

Prompt 2:

I am a construction worker. I am interested in building a house. Is there any way to use polymer composites for 3D printing?

| Response 1 | Response 2 |
|---|---|
| Yes, polymer composites can be used for 3D printing in construction. This technology, known as additive manufacturing, is gaining traction in the construction industry for building houses and other structures due to its efficiency, cost-effectiveness, and design flexibility. Here are some key points and resources regarding the use of polymer composites in 3D printing for construction: <br><br> 1. Advantages of Using Polymer Composites in 3D Printing <br> • Material Properties: Polymer composites offer high strength-to-weight ratios, durability, and resistance to environmental factors. <br> • Design Flexibility: 3D printing allows for complex and customized designs that are difficult to achieve with traditional construction methods. <br> • Sustainability: Using recycled materials and optimizing material use can reduce waste and environmental impact. <br><br> 2. Applications in Construction <br> • Walls and Structural Components: Polymer composites can be used to print walls, partitions, and other structural components, offering enhanced insulation and strength. <br> • Decorative Elements: Complex architectural details and decorative elements can be easily printed with polymer composites. <br><br> 3. Technologies and Techniques <br> • Large-Scale 3D Printing: Techniques like Big Area Additive Manufacturing (BAAM) can print large parts and structures using polymer composites. This method is suitable for constructing entire sections of buildings. <br> • Hybrid Processes: Combining additive manufacturing with traditional construction methods can optimize the use of polymer composites in building construction. <br><br> 4. Examples and Case Studies <br> • MDF's Work on Additive Manufacturing: The Manufacturing Demonstration Facility (MDF) at ORNL has developed innovative techniques for large-scale additive manufacturing using polymer composites. This includes creating structural components and entire sections of buildings. <br><br> 5. Expert Contacts for Further Guidance <br> Connecting with experts in the field can provide you with the knowledge and resources needed to implement these technologies in your construction projects: <br><br> Dr. Ahmed A. Hassen <br> Group Leader & Senior R&D Staff Scientist, Composites Innovation Group, ORNL Email: hassenaa@ornl.gov <br> Dr. Hassen specializes in high-rate manufacturing of advanced composite structures and developing strategies for smart manufacturing of polymers and advanced composites. <br><br> John M. Lindahl <br> Technical Associate Staff, ORNL Email: lindahljm1@ornl.gov <br> John Lindahl has extensive experience in large-scale polymer additive manufacturing and can provide insights into the practical applications of this technology in construction. <br><br> 6. Further Reading and Collaboration Opportunities <br> • ORNL's Tech Collaboration Program: Engages industry partners in projects to develop and deploy new manufacturing processes and materials. This program can provide valuable insights and resources for your construction projects.- Publications and Case Studies: Review publications and case studies from ORNL and other research institutions to understand the latest advancements and best practices in using polymer composites for 3D printing in construction. <br><br> By leveraging these resources and technologies, you can effectively incorporate polymer composites into your construction projects, enhancing the efficiency and quality of your builds | Using polymer composites for 3D printing in construction is an emerging field with significant potential. Here are some ways polymer composites can be used in 3D printing for building a house, along with relevant resources and steps to get started: <br><br> 1. Understanding the Basics of 3D Printing in Construction <br> • 3D Printing Techniques: Common methods include Fused Deposition Modeling (FDM), Stereolithography (SLA), and Selective Laser Sintering (SLS). For construction, large-scale FDM and Concrete Printing are most relevant. <br> • Materials: Polymer composites used in 3D printing typically consist of a polymer matrix (like PLA, ABS, or nylon) reinforced with fibers (such as carbon, glass, or aramid). <br><br> 2. Advantages of Using Polymer Composites in Construction <br> • Strength and Durability: Polymer composites can provide enhanced structural integrity. <br> • Lightweight: Reduced weight compared to traditional materials, which can lower transportation and handling costs. <br> • Design Flexibility: Complex shapes and structures that are difficult to achieve with traditional methods. <br> • Thermal and Acoustic Insulation: Improved insulation properties compared to conventional materials. <br><br> 3. 3D Printing Technologies for Construction <br> • Large-Scale FDM: Companies like WASP and CyBe Construction use large-scale FDM printers capable of printing entire structures. <br> • Concrete Printing: This involves extruding a composite material made of cementitious components combined with polymers to enhance properties. <br><br> 4. Key Companies and Technologies <br> • ICON: Specializes in 3D printing concrete homes. <br> • Apis Cor: Known for their 3D printed houses using a mobile printer. <br> • Contour Crafting: A method developed by Dr. Behrokh Khoshnevis for large-scale 3D printing of buildings. <br><br> 5. Materials and Composites for 3D Printing in Construction <br> • Polymer Concrete: A composite material where polymers replace cement as a binder, providing better resistance to corrosion and chemical attacks. <br> • Fiber-Reinforced Polymers (FRPs): Used to enhance the mechanical properties of the printed structures. <br><br> 6. Steps to Get Started <br> • Research and Learning <br> • Courses and Tutorials: <br> • On.line platforms like Coursera, edX, and Udacity offer courses on 3D printing and materials science. <br> • Specialized courses on construction 3D printing by institutions like the Eindhoven University of Technology. <br> • Books: <br> • "Additive Manufacturing: Materials, Processes, Quantifications and Applications" by Kun Zhou. <br> • "3D Printing in Construction: A Practical Guide" by Berenika Hausnerova. <br><br> Hands-On Experience <br> • Workshops: Participate in workshops and training sessions focused on 3D printing and construction. <br> • DIV Projects: Start with smaller projects to understand the material properties and printing techniques. <br><br> Setting Up Your 3D Printing System <br> • Printer Selection: Choose a printer suitable for construction purposes. For large-scale projects, look into machines from companies like WASP, ICON, and Apis Cor. <br> • Material Sourcing: Source appropriate polymer composites, considering factors like strength, durability, and environmental resistance. <br><br> Building and Experimentation <br> • Prototyping: Begin with small-scale models to test the design and material properties. <br> • Full-Scale Projects: Once comfortable, move to larger projects, ensuring you follow building codes and safety regulations. <br><br> 7. Networking and Community <br> • Forums and Groups: Join online forums and groups focused on 3D printing and construction, like those on LinkedIn and Reddit. <br> • Conferences and Trade Shows: Attend events like the International Conference on 3D Printing in Construction to network with industry professionals and learn about the latest developments. <br><br> Resources and References <br> • Websites and Biogs: Follow websites like 3DPrint.com and Construction3DPrinting.com for the latest news and advancements. <br> • Professional Organizations: Engage with organizations such as the American Composites Manufacturers Association (ACMA) and the Society of Manufacturing Engineers (SME). <br><br> By leveraging these resources and steps, you can explore and potentially implement polymer composites for 3D printing in your construction projects, bringing innovative and efficient methods to building houses. |

Fig. 5 User Study 1 Prompt 2. The second prompt out of six total prompts and corresponding responses provided to users during the first user study. Response 1 was generated by the Composites Guide Model and Response 2 was generated by the standard GPT-4o model.

Prompt 3:

I want to fabricate an ID card holder out of composites. How would I do that?

| Response 1 | Response 2 |
|---|---|
| Fabricating an ID card holder out of composites involves several steps, from design to material selection and manufacturing. Here's a comprehensive guide to help you get started:<br><br>1. Design the ID Card Holder<br>• CAD Software: Use computer-aided design (CAD) software like AutoCAD, SolidWorks, or Fusion 360 to design your ID card holder. Ensure the design includes dimensions and features like slots for the card and attachment points for lanyards or clips.<br><br>2. Select Materials<br>• Polymer Matrix: Common options include epoxy resin, polyester resin, or thermoplastics like ABS.<br>• Reinforcement Fibers: Choose from glass fiber, carbon fiber, or Kevlar. Carbon fiber is strong and lightweight, making it a good choice for small items like ID card holders.<br><br>3. Prepare the Mold<br>• Material for Mold: For a small item like an ID card holder, you can use materials like silicone, aluminum, or 3D printed plastic for the mold.<br>• Mold Design: Ensure the mold has the correct dimensions and includes any features you designed in the CAD model.<br><br>4. Fabrication Methods<br>• Hand Lay-Up: Simple and cost-effective for small-scale production.<br>• Steps:<br>• 1. Apply a release agent to the mold to prevent sticking.<br>• 2. Lay the reinforcement fibers in the mold.<br>• 3. Apply the resin mixture to the fibers.<br>• 4. Use a roller to remove air bubbles and ensure even distribution of the resin.<br>• 5. Allow the resin to cure as per the manufacturer's instructions.<br>• Resin Transfer Molding (RTM): Suitable for higher precision and better surface finish.<br>• Steps:<br>• 1. Place dry fibers in the mold.<br>• 2. Close the mold and inject resin under pressure.<br>• 3. Allow the resin to cure.<br>• Vacuum Bagging: Improves the fiber-to-resin ratio and removes air bubbles.<br>• Steps:<br>• 1. Lay the fibers and apply resin in the mold.<br>• 2. Cover with a peel ply, breather fabric, and vacuum bagging film.<br>• 3. Apply vacuum pressure to remove excess resin and air.<br>• 4. Allow to cure under vacuum pressure.<br><br>5. Curing<br>• Room Temperature: Some resins cure at room temperature over several hours.<br>• Heat Curing: For faster curing, use an oven or heat lamps as specified by the resin manufacturer.<br><br>6. Finishing<br>• Trimming: Trim excess material using a rotary tool or fine saw.<br>• Sanding: Sand the edges to achieve a smooth finish.<br>• Polishing: Apply a polishing compound for a glossy finish if desired.<br><br>7. Quality Control<br>• Inspection: Check for any defects, such as air bubbles, incomplete curing, or irregularities in the shape.<br>• Testing: Ensure the ID card fits correctly and the holder functions as intended.<br><br>8. Customization and Assembly<br>• Painting and Coating: Apply paint or protective coatings if required.<br>• Assembly: Attach any additional components like clips, lanyard loops, or magnetic strips.<br><br>Tools and Materials List<br>• CAD Software<br>• Release Agent<br>• Reinforcement Fibers (e.g., carbon fiber)<br>• Resin (e.g., epoxy)<br>• Mold Material (e.g., silicone or aluminum)<br>• Vacuum Bagging Kit (if using vacuum bagging)<br>• Rotary Tool/Saw<br>• Sandpaper<br>• Polishing Compound<br><br>Additional Resources<br>• Videos and Tutorials: Search for videos on YouTube for visual guides on composite lay-up and mold making.<br>• Books: "Composite Materials: Fabrication Handbook" by John Wanberg provides practical insights into working with composites.<br>• Online Courses: Platforms like Coursera and Udemy offer courses on composite materials and fabrication techniques.<br><br>By following these steps, you can fabricate a durable and lightweight ID card holder out of composites, leveraging the strength and versatility of composite materials. | Fabricating an ID card holder using polymer composites involves several steps, including material selection, design, molding, and finishing. Here's a detailed guide on how you can do this:<br><br>1. Material Selection<br>Choose appropriate polymer composite materials based on the desired properties of the ID card holder, such as strength, flexibility, and aesthetics. Common materials include:<br>• Fiberglass Reinforced Plastic (FRP): Offers good strength and durability.<br>• Carbon Fiber Reinforced Polymer (CFRP): Provides high strength-to-weight ratio and rigidity.<br>• Thermoplastic Composites: These are easier to mold and can be recycled.<br><br>2. Design<br>Create a design for the ID card holder using CAD (Computer-Aided Design) software. Ensure that the design includes the necessary dimensions and features, such as slots for the card and attachment points for lanyards or clips.<br><br>3. Mold Preparation<br>If you are using molding techniques, you will need a mold. This can be made from metal, silicone, or even 3D-printed plastic, depending on the chosen manufacturing process.<br><br>4. Fabrication Process<br>Here are a few methods you can use to fabricate the ID card holder:<br><br>Injection Molding<br>1. Prepare the Mold: Clean and prepare the mold.<br>2. Melt the Composite Material: Heat the thermoplastic composite material until it reaches a molten state.<br>3. Inject the Material: Inject the molten material into the mold under high pressure.<br>4. Cooling and Ejection: Allow the material to cool and solidify before ejecting the part from the mold.<br><br>Compression Molding<br>1. Prepare the Mold: Clean and preheat the mold.<br>2. Place the Composite Material: Place a pre-measured amount of composite material into the mold cavity.<br>3. Close the Mold: Close the mold and apply heat and pressure to shape the material.<br>4. Cooling and Removal: Allow the mold to cool before opening it and removing the finished part.<br><br>3D Printing<br>1. Prepare the Design: Convert the CAD design into a format compatible with the 3D printer.<br>2. Set Up the Printer: Load the composite filament (e.g., a thermoplastic composite filament) into the 3D printer.<br>3. Print the Part: Print the ID card holder layer by layer according to the design.<br>4. Post-Processing: Remove any support structures and smooth the surface if necessary.<br><br>5. Finishing<br>After fabrication, you may need to perform some finishing steps:<br>• Trimming: Remove any excess material or flash.<br>• Sanding: Smooth out rough edges and surfaces.<br>• Painting or Coating: Apply any desired finishes, such as paint or protective coatings.<br><br>6. Quality Control<br>Inspect the final product for any defects or inconsistencies. Ensure that all dimensions are accurate and that the ID card fits properly.<br><br>Expert Contacts<br>For more detailed guidance and support, consider reaching out to experts in composite manufacturing:<br><br>Dr. Ahmed A. Hassen<br>Group Leader & Senior R&D Staff Scientist, Composites Innovation Group, ORNL<br>Email: hassenaa@ornl.gov<br>Dr. Hassen specializes in high-rate manufacturing of advanced composite structures and can provide insights into composite material selection and processing techniques.<br><br>John M. Lindahl<br>Technical Associate Staff, ORNL Email: lindahljm1@ornl.gov<br>John Lindahl has extensive experience in large-scale polymer additive manufacturing and can offer practical advice on the fabrication process.<br><br>By following these steps and utilizing available resources, you can successfully fabricate a durable and aesthetically pleasing ID card holder out of composite materials. |

*Fig. 6 User Study 1 Prompt 3. The third prompt out of six total prompts and corresponding responses provided to users during the first user study. Response 1 was generated by the standard GPT-4o model and Response 2 was generated by the Composites Guide model.*

Prompt 4:

What is the best way to fabricate a side-view mirror of a car?

| Response 1 | Response 2 |
|---|---|
| Fabricating a side-view mirror of a car using polymer composites involves several advanced techniques to ensure the final product is strong, lightweight, and has the desired properties. Here is a comprehensive guide on how to achieve this:<br><br>1. Material Selection<br>Choosing the right materials is crucial for the durability and performance of the side-view mirror. Common materials include:<br>• Carbon Fiber Reinforced Polymer (CFRP): Offers high strength-to-weight ratio and excellent rigidity.<br>• Glass Fiber Reinforced Polymer (GFRP): Provides good strength and cost-effectiveness.<br>• Thermoplastic Composites: These can be easily molded and recycled.<br><br>2. Design<br>Create a detailed design using CAD software, considering the aerodynamic properties, mounting features, and aesthetics. The design should include:<br>• The mirror housing<br>• The mounting bracket<br>• The reflective surface and its attachment mechanism<br><br>3. Fabrication Process<br>Depending on the chosen materials and design complexity, several fabrication methods can be employed:<br><br>1. Injection Molding (for Thermoplastic Composites)<br>• Design the Mold: Create a mold that matches the CAD design.<br>• Melt the Composite Material: Use an injection molding machine to melt the thermoplastic composite.<br>• Inject the Material: Inject the molten material into the mold under high pressure.<br>• Cooling and Ejection: Allow the material to cool and solidify before removing it from the mold.<br><br>2. Compression Molding (for Thermoset and Thermoplastic Composites)<br>• Prepare the Mold: Clean and preheat the mold.<br>• Place the Composite Material: Place the pre-measured composite material into the mold cavity.'<br>• Close the Mold: Apply heat and pressure to shape the material.<br>• Cooling and Removal: Cool the mold before opening it and removing the finished part.<br><br>3. Resin Transfer Molding (RTM) (for Thermoset Composites)<br>• Create a Mold: Design a mold for the mirror housing.<br>• Lay the Fibers: Place carbon or glass fibers in the mold.<br>• Inject the Resin: Inject resin into the mold to impregnate the fibers.<br>• Curing: Allow the resin to cure, forming a solid composite part.<br>• Demolding: Remove the finished part from the mold.<br><br>4. Additive Manufacturing (3D Printing)<br>• Prepare the Design: Convert the CAD design into a format compatible with the 3D printer.<br>• Set Up the Printer: Load the composite filament (e.g., carbon fiber-reinforced thermoplastic) into the 3D printer.<br>• Print the Part: Print the side-view mirror housing layer by layer.<br>• Post-Processing: Remove support structures and smooth the surface if necessary.<br><br>4. Assembly<br>• Attach the Reflective Surface: Secure the mirror glass to the housing using adhesive or mechanical fasteners.<br>• Install the Mounting Bracket: Attach the bracket to the mirror housing.<br>• Finish and Paint: Apply any desired finishes or paint to the mirror housing to match the car's exterior.<br><br>5. Quality Control<br>Inspect the final product for defects, ensure proper fitment, and test for durability and performance.<br><br>Expert Contacts<br><br>For further assistance and detailed technical guidance, consider reaching out to the following experts:<br><br>Dr. Ahmed A. Hassen<br>Group Leader & Senior R&D Staff Scientist, Composites Innovation Group, ORNL Email: hassenaa@ornl.gov<br>Dr. Hassen has extensive experience in high-rate manufacturing of advanced composite structures and can provide valuable insights into the fabrication process.<br><br>John M. Lindahl<br>Technical Associate Staff, ORNL Email: lindahljm1@ornl.gov<br>John Lindahl's expertise in large-scale polymer additive manufacturing can offer practical advice on 3D printing techniques and material selection.<br><br>By following these steps and utilizing the available resources, you can fabricate a high-quality side- view mirror for a car using polymer composites. | Fabricating a side-view mirror for a car using composite materials involves several detailed steps to ensure durability, aerodynamic efficiency, and aesthetics. Here's a comprehensive guide on how to do it:<br><br>1. Design and Prototyping<br>• CAD Modeling: Use CAD software like SolidWorks, AutoCAD, or Fusion 360 to design the side-view mirror. Ensure the design accommodates the mirror glass, mounting points, and any electronic components (e.g., turn signal lights, heating elements).<br>• Prototype: Create a 3D-printed prototype to verify the design, fit, and functionality before moving to composite fabrication.<br><br>2. Material Selection<br>• Polymer Matrix: Epoxy resin is commonly used for its excellent mechanical properties and resistance to environmental factors.<br>• Reinforcement Fibers: Carbon fiber is preferred for automotive applications due to its high strength-to-weight ratio and stiffness. Glass fiber is a more cost-effective alternative but with lower strength and stiffness.<br><br>3. Mold Preparation<br>• Mold Design: Design the mold based on your CAD model. The mold can be split into two halves to facilitate easy removal of the finished part.<br>• Mold Material: Use materials like aluminum for a durable and reusable mold or high-quality 3D- printed plastic for a cost-effective option.<br>• Surface Finish: Ensure the mold has a smooth surface finish to achieve a high-quality finish on the composite part. Apply a release agent to the mold to prevent sticking.<br><br>4. Fabrication Methods<br>• Hand Lay-Up and Vacuum Bagging: This method ensures good fiber-to-resin ratios and eliminates air bubbles.<br>• Steps:<br>• 1. Cutting Fibers: Cut carbon fiber cloth to fit the mold, considering the lay-up sequence for optimal strength.<br>• 2. Lay-Up: Apply layers of carbon fiber cloth into the mold, brushing epoxy resin onto each layer.<br>• 3. Vacuum Bagging: Cover the lay-up with a peel ply, breather fabric, and vacuum bagging film. Apply vacuum pressure to remove excess resin and air.<br>• 4. Curing: Allow the resin to cure under vacuum pressure. Follow the resin manufacturer's instructions for curing times and temperatures.<br>• Resin Transfer Molding (RTM): Suitable for high-quality, complex shapes, and higher production volumes.<br>• Steps:<br>• 1. Dry Lay-Up: Place dry carbon fiber preforms in the mold.<br>• 2. Close Mold and Inject Resin: Close the mold and inject epoxy resin under pressure, ensuring complete wet-out of the fibers.<br>• 3. Curing: Allow the resin to cure under controlled conditions.<br><br>5. Post-Processing<br>• Trimming: Trim excess material using a rotary tool or fine saw.<br>• Sanding: Sand the edges and surface to achieve a smooth finish.<br>• Painting and Coating: Apply a primer, paint, and clear coat for aesthetics and UV protection if required. Use automotive-grade paints and coatings.<br><br>6. Assembly<br>• Mirror Glass: Cut and fit the mirror glass into the composite housing. Secure it with appropriate adhesives or mounting brackets.<br>• Electronic Components: Install any electronic components such as turn signals, heating elements, or motors.<br>• Mounting: Ensure the side-view mirror assembly has secure mounting points compatible with the vehicle.<br><br>7. Quality Control<br>• Inspection: Check for any defects, such as air bubbles, incomplete curing, or surface imperfections.<br>• Testing: Ensure the side-view mirror fits correctly, functions as intended, and meets safety and aerodynamic standards.<br><br>Tools and Materials List<br>• CAD Software<br>• Epoxy Resin<br>• Carbon Fiber Cloth<br>• Mold Material (Aluminum or 3D-Printed Plastic)<br>• Release Agent<br>• Vacuum Bagging Kit<br>• Rotary Tool/Saw<br>• Sandpaper<br>• Automotive-Grade Paint and Clear Coat<br>• Mirror Glass<br>• Electronic Components (optional)<br><br>Additional Resources<br>• Videos and Tutorials: YouTube channels like "Easy Composites" provide detailed tutorials on working with carbon fiber and vacuum bagging.<br>• Books: "Composite Materials: Fabrication Handbook" by John Wanberg offers practical insights into composite fabrication techniques.<br>• Online Courses: Platforms like Udemy and Coursera offer courses on composite materials and manufacturing processes.<br><br>By following these steps, you can fabricate a high-quality side-view mirror using composite materials, taking advantage of their strength, lightweight properties, and aesthetic flexibility. |

Fig. 7 User Study 1 Prompt 4. The fourth prompt out of six total prompts and corresponding responses provided to users during the first user study. Response 1 was generated by the Composites Guide and Response 2 was generated by the standard GPT-4o model.

Prompt 5:

How would I design and create a lightweight automotive frame?

| Response 1 | Response 2 |
|---|---|
| Designing and creating a lightweight automotive frame involves several key steps, from conceptual design to material selection and manufacturing processes. Here's a comprehensive guide to help you through the process:<br><br>1. Conceptual Design and Requirements<br>• Define Objectives: Determine the key objectives for the frame, such as weight reduction, strength, stiffness, safety, cost, and manufacturability.<br>• Regulatory Compliance: Ensure the design meets automotive safety and performance standards, such as those set by the National Highway Traffic Safety Administration (NHTSA) and the European New Car Assessment Programme (Euro NCAP).<br>2. Material Selection<br>• Lightweight Materials:<br>• Aluminum Alloys: Good strength-to-weight ratio and corrosion resistance.<br>• High-Strength Steel (HSS): Provides high strength and durability.<br>• Carbon Fiber-Reinforced Polymers (CFRP): Extremely lightweight and strong but more expensive.<br>• Magnesium Alloys: Lightweight but less commonly used due to cost and fabrication challenges.<br>• Hybrid Solutions: Combining different materials to optimize performance and cost (e.g., aluminum frame with carbon fiber reinforcements).<br>3. Design Process<br>• CAD Modeling: Use CAD software (e.g., SolidWorks, CATIA, or Autodesk Inventor) to create detailed 3D models of the frame. Focus on:<br>• Structural integrity<br>• Integration with other vehicle components<br>• Finite Element Analysis (FEA): Perform FEA to simulate and analyze the frame's structural performance under various loading conditions. Tools like ANSYS or Abaqus are commonly used.<br>• Topology Optimization: Use optimization algorithms to reduce weight by removing unnecessary material while maintaining strength and stiffness.<br>4. Prototyping and Testing<br>• Rapid Prototyping: Create scale models or specific sections using 3D printing to quickly evaluate design concepts.<br>• Physical Prototyping: Fabricate a full-scale prototype of the frame. This can be done using CNC machining, welding, and composite lay-up techniques, depending on the chosen materials.<br>• Testing: Conduct physical tests, including static load tests, dynamic impact tests, and fatigue tests to validate the design.<br>5. Manufacturing Process<br>• Aluminum and Steel Frames:<br>• Extrusion: For creating complex cross-sectional profiles.<br>• Stamping and Forming: For sheet metal components.<br>• Welding: Techniques like MIG, TIG, and resistance welding for joining.<br>• Composite Frames:<br>• Lay-Up and Vacuum Bagging: Manual process for placing and curing composite layers.<br>• Resin Transfer Molding (RTM): Automated process for higher volume production.<br>• Autoclave Curing: For high-performance components requiring precise temperature and pressure control.<br>• Hybrid Frames: Combining the above techniques for different sections of the frame.<br>6. Quality Control and Inspection<br>• Dimensional Inspection: Ensure all dimensions are within tolerances using tools like CMM (Coordinate Measuring Machine).<br>• Non-Destructive Testing (NDT): Methods like ultrasonic testing, X-ray, and dye penetrant inspection to check for internal defects.<br>• Performance Testing: Reiterate tests such as crash tests, torsional rigidity tests, and durability tests to ensure the frame meets all performance criteria.<br>7. Integration and Assembly<br>• Component Integration: Ensure the frame is designed to integrate seamlessly with suspension systems, powertrain, body panels, and other vehicle components.<br>• Modular Design: Consider a modular approach to facilitate easier assembly and maintenance.<br><br>Tools and Resources<br>• CAD Software: SolidWorks, CATIA, Autodesk Inventor<br>• FEA Software: ANSYS, Abaqus<br>• Optimization Software: OptiStruct, Tosca<br>• Manufacturing Equipment: CNC machines, welding equipment, autoclaves, composite lay-up tools<br><br>Additional Resources<br>• Books:<br>• "Automotive Chassis Engineering" by David C. Barton<br>• "Lightweight Materials: Understanding the Basics" by V. S. Aigbodion and S. B. Hassan<br>• Online Courses: Platforms like Coursera, edX, and Udemy offer courses on automotive engineering, materials science, and CAD/FEA techniques.<br>• Professional Organizations: Join organizations like SAE International for access to technical papers, standards, and industry conferences.<br><br>By following these steps and utilizing the recommended tools and resources, you can design and create a lightweight automotive frame that meets your objectives for performance, safety, and efficiency. | Designing and creating a lightweight automotive frame using polymer composites involves several critical steps to ensure the structure is strong, durable, and meets all safety standards. Here's a comprehensive guide to help you through the process:<br><br>1. Material Selection<br>Choose appropriate composite materials based on the required properties, such as high strength-to-weight ratio, corrosion resistance, and impact resistance. Common materials include:<br>• Carbon Fiber Reinforced Polymer (CFRP): Offers excellent strength-to-weight ratio and rigidity.<br>• Glass Fiber Reinforced Polymer (GFRP): Provides good strength and is cost-effective.<br>• Hybrid Composites: Combining different types of fibers (e.g., carbon and glass) to optimize performance and cost.<br>2. Design Process<br>Use Computer-Aided Design (CAD) software to create a detailed design of the automotive frame. Consider the following:<br>• Structural Requirements: Ensure the design meets safety and performance standards.<br>• Weight Optimization: Minimize weight without compromising strength.<br>• Integration of Components: Design the frame to accommodate all necessary components (e.g., engine, transmission, suspension).<br>3. Simulation and Analysis<br>Perform Finite Element Analysis (FEA) to simulate and analyze the performance of the frame under various conditions. This helps in:<br>• Identifying Stress Points: Determine areas of high stress and potential failure.<br>• Optimizing Material Distribution: Ensure materials are used efficiently to enhance strength and reduce weight.<br>4. Fabrication Methods<br>Several fabrication methods can be used depending on the design complexity and material choice:<br>• 1. Resin Transfer Molding (RTM)<br>• Create a Mold: Design and manufacture a mold based on the CAD model.<br>• Lay the Fibers: Place carbon or glass fibers into the mold.<br>• Inject the Resin: Inject resin into the mold to impregnate the fibers.<br>• Curing: Allow the resin to cure, forming a solid composite structure.<br>• Demolding: Remove the finished frame from the mold.<br>• 2. Filament Winding<br>• Mandrel Preparation: Create a mandrel in the shape of the frame components.<br>• Winding Process: Wind continuous fiber filaments around the mandrel.<br>• Resin Application: Apply resin to the wound fibers.<br>• Curing: Cure the resin to solidify the composite material.<br>• Mandrel Removal: Remove the mandrel, leaving the composite frame component.<br>• 3. Additive Manufacturing (3D Printing)<br>• Prepare the Design: Convert the CAD design into a format compatible with the 3D printer.<br>• Set Up the Printer: Load composite filament (e.g., carbon fiber-reinforced thermoplastic) into the 3D printer.<br>• Print the Frame: Print the frame components layer by layer.<br>• Post-Processing: Remove support structures and smooth the surfaces if necessary.<br>5. Assembly<br>• Join Components: Use adhesive bonding, mechanical fasteners, or welding (for metal-composite hybrid structures) to assemble the frame components.<br>• Integration of Subsystems: Attach other automotive subsystems (engine, transmission, suspension) to the frame.<br>6. Testing and Validation<br>Conduct rigorous testing to ensure the frame meets all safety and performance standards:<br>• Crash Testing: Assess the frame's impact resistance and safety.<br>• Fatigue Testing: Evaluate the frame's durability under repeated loading.<br>• Environmental Testing: Test the frame's resistance to environmental factors (e.g., corrosion, temperature changes).<br><br>Expert Contacts<br>For more detailed guidance and technical support, consider reaching out to the following experts:<br><br>Dr. Ahmed A. Hassen<br>Group Leader & Senior R&D Staff Scientist, Composites Innovation Group,<br>ORNL Email: hassenaa@ornl.gov<br><br>Dr. Hassen specializes in high-rate manufacturing of advanced composite structures and can provide insights into material selection and fabrication techniques.<br><br>Dr. Cait M. Clarkson<br>Associate R&D Staff Scientist, Sustainable Manufacturing Technologies Group, ORNL Email: clarksoncm@ornl.gov<br>Dr. Clarkson focuses on sustainable material feedstocks and their integration into advanced manufacturing processes, which can be beneficial for designing lightweight and eco-friendly automotive frames.<br><br>By following these steps and leveraging the expertise available, you can design and create a lightweight automotive frame that meets all necessary performance and safety requirements. |

Fig. 8 User Study 1 Prompt 5. The fifth prompt out of six total prompts and corresponding responses provided to users during the first user study. Response 1 was generated by the standard GPT-4o model and Response 2 was generated by the Composites Guide model.

Prompt 6:

I am trying to make a glass-fiber-reinforced polymer composite panel for the car hood with thermoset materials. It should have high impact resistance and high temperature resistance. Please suggest which fiber layups and resin materials are best as well as manufacturing processes.

| Response 1 | Response 2 |
|---|---|
| Creating a glass-fiber-reinforced polymer (GFRP) composite panel for a car hood with high impact resistance and high temperature resistance involves careful selection of fibers, resins, and manufacturing processes. Here's a detailed guide:<br>1. Material Selection Fibers<br>For high impact resistance and temperature resistance, use high-performance glass fibers such as S-Glass or E-Glass:<br>• S-Glass: Higher strength and stiffness compared to E-Glass, suitable for high-performance applications.<br>• E-Glass: Standard glass fiber with good overall performance and cost-effectiveness.<br><br>Resins<br>Thermoset resins with high temperature and impact resistance are crucial. Consider the following options:<br>• Epoxy Resin: Offers excellent mechanical properties, high temperature resistance, and good adhesion to fibers.<br>• Vinyl Ester Resin: Provides good resistance to heat and chemicals, and has better impact resistance than polyester resins.<br>• Phenolic Resin: Excellent flame resistance and high-temperature performance, though it can be more brittle.<br><br>2. Fiber Layup<br>The layup pattern significantly affects the mechanical properties of the composite. Consider the following configurations:<br>• Quasi-Isotropic Layup: This involves stacking layers in multiple orientations (e.g., 0°, 45°, -45°, 90°) to provide balanced properties in all directions.<br>• Cross-Ply Layup: Alternating layers of 0° and 90° orientations provide good stiffness and strength in two directions.<br>• Hybrid Layup: Combine different types of fibers or layup patterns to optimize performance.<br><br>A common approach for high-performance applications is a combination of unidirectional and woven fabrics:<br>• Unidirectional (UD) Fibers: For maximum strength and stiffness in specific directions.<br>• Woven Fabrics: For improved impact resistance and dimensional stability.<br><br>3. Manufacturing Processes Resin Transfer Molding (RTM)<br>1. Mold Preparation: Clean and apply release agent to the mold.<br>2. Fiber Layup: Place pre-cut glass fiber fabrics in the mold according to the chosen layup pattern.<br>3. Resin Injection: Inject the thermoset resin into the mold to impregnate the fibers.<br>4. Curing: Allow the resin to cure at elevated temperatures, typically in an oven or using heated molds.<br>5. Demolding: Remove the cured composite panel from the mold.<br><br>Vacuum-Assisted Resin Transfer Molding (VARTM)<br>1. Mold Preparation: Apply release agent and position the fiber layup in the mold.<br>2. Vacuum Bagging: Cover the mold with a vacuum bag and seal it.<br>3. Resin Infusion: Apply vacuum to draw the resin into the mold, ensuring thorough impregnation of the fibers.<br>4. Curing: Cure the composite under vacuum at elevated temperatures.<br>5. Demolding: Remove the composite panel from the mold.<br><br>Hand Lay-Up with Vacuum Bagging<br>1. Mold Preparation: Apply release agent to the mold.<br>2. Fiber Layup: Manually lay the glass fiber fabrics in the mold and apply resin to each layer using brushes or rollers.<br>3. Vacuum Bagging: Cover the layup with a vacuum bag and apply vacuum to remove air bubbles and excess resin.<br>4. Curing: Allow the composite to cure under vacuum at room temperature or elevated temperatures.<br>5. Demolding: Remove the cured composite panel from the mold.<br><br>Recommended Materials and Processes<br>• Fibers: S-Glass or E-Glass with a quasi-isotropic or cross-ply layup.<br>• Resin: Epoxy resin for the best balance of mechanical properties and temperature resistance.<br>• Process: Resin Transfer Molding (RTM) or Vacuum-Assisted Resin Transfer Molding (VARTM) for high-quality, high-performance parts.<br><br>Expert Contacts<br><br>For further assistance and detailed guidance, consider reaching out to the following experts:<br><br>Dr. Ahmed A. Hassen<br>Group Leader & Senior R&D Staff Scientist, Composites Innovation Group, ORNL Email: hassenaa@ornl.gov<br>Dr. Hassen specializes in high-rate manufacturing of advanced composite structures and can provide insights into material selection and fabrication techniques.<br><br>By following these guidelines, you can create a high-performance glass-fiber-reinforced polymer composite panel for a car hood that meets your requirements for impact resistance and temperature resistance. | Creating a glass-fiber-reinforced polymer composite panel for a car hood involves selecting the right materials and manufacturing processes to ensure high impact resistance and high temperature resistance. Here's a comprehensive guide on fiber layups, resin materials, and manufacturing processes that would be suitable for your application:<br>Material Selection Fiber Layups<br>• Woven Roving: Provides high strength and is suitable for structural applications. Woven roving consists of continuous fibers woven into a fabric, offering excellent impact resistance.<br>• Unidirectional (UD) Fibers: Ideal for areas requiring high tensile strength in a specific direction. These can be combined with woven roving for optimized performance.<br>• Multiaxial Fabrics: Includes biaxial (±45°), triaxial (0°/±45°), and quadraxial (0°/±45°/90°) fabrics. These offer strength in multiple directions and are great for complex load conditions.<br><br>Resin Materials<br>• Epoxy Resin: Known for its excellent mechanical properties, adhesion, chemical resistance, and high-temperature performance. Epoxy resins are commonly used in automotive and aerospace applications.<br>• Vinyl Ester Resin: Offers good mechanical properties and superior chemical resistance compared to polyester resins. It also has good heat resistance, making it suitable for high-temperature applications.<br>• Polyester Resin: While cost-effective, it has lower mechanical properties and heat resistance compared to epoxy and vinyl ester resins. It's less ideal for high-performance applications but can be considered for cost-sensitive projects.<br>Manufacturing Processes Vacuum Bagging<br>• Process: Involves laying up the composite layers in a mold, then covering with a vacuum bag and applying vacuum pressure. This process removes air voids and excess resin, leading to a high-quality laminate.<br>• Advantages: Good fiber-to-resin ratio, improved laminate quality, and relatively simple setup.<br>• Suitable for: Small to medium-sized parts and prototypes.<br><br>Resin Transfer Molding (RTM)<br>• Process: Involves placing dry fiber preforms into a closed mold and injecting resin under pressure. The resin impregnates the fibers, and the part cures within the mold.<br>• Advantages: Produces high-quality, complex parts with good surface finish and repeatability. Suitable for medium to high production volumes.<br>• Suitable for: Complex geometries and higher production volumes.<br><br>Hand Lay-Up<br>• Process: The simplest and most manual method, where fibers are manually placed in the mold and resin is applied with brushes or rollers.<br>• Advantages: Low cost and simple setup, suitable for prototypes and low-volume production.<br>• Disadvantages: Lower quality control, higher risk of air voids, and inconsistent fiber-to-resin ratio.<br>Recommended Approach Fiber Layup<br>• Hybrid Layup: Use a combination of woven roving for overall strength and unidirectional fibers in high-stress areas. Consider a sandwich construction with a core material (e.g., foam or honeycomb) to increase stiffness without significantly increasing weight.<br>• Example Layup:<br>• Outer layer: Woven roving (for impact resistance and surface finish)<br>• Middle layers: Unidirectional fibers oriented in different directions (for strength in multiple directions)<br>• Inner layer: Woven roving or another layer of UD fibers<br>• Optional: Core material for added stiffness and reduced weight<br><br>Resin Selection<br>• Primary Choice: Epoxy resin for its superior mechanical properties, impact resistance, and high-temperature performance.<br>• Alternative Choice: Vinyl ester resin if cost is a significant concern and slightly lower performance can be tolerated.<br><br>Manufacturing Process<br>• Preferred Method: Vacuum bagging for its balance between quality and cost-effectiveness, especially for custom or low-volume production.<br>• Alternative Method: Resin Transfer Molding (RTM) if producing medium to high volumes or if superior surface finish and part consistency are required.<br><br>Steps for Fabrication<br><br>1. Design and Mold Preparation<br>• Create a detailed CAD model of the car hood.<br>• Manufacture a high-quality mold from materials such as aluminum or fiberglass.<br><br>2. Layup Process<br>• Apply a release agent to the mold.<br>• Lay the first layer of woven roving, ensuring it conforms to the mold shape.<br>• Apply epoxy resin to the first layer.<br>• Add layers of unidirectional fibers in the desired orientations, applying resin between each layer.<br>• Repeat with woven roving or other selected fabrics.<br><br>3. Vacuum Bagging<br>• Cover the layup with peel ply, breather fabric, and vacuum bagging film.<br>• Apply vacuum pressure to remove air voids and excess resin.<br>• Allow the part to cure under vacuum pressure, following the epoxy resin manufacturer's curing schedule.<br><br>4. Post-Processing<br>• Trim excess material and sand the edges for a smooth finish.<br>• Conduct any necessary surface finishing, such as painting or coating.<br><br>5. Quality Control<br>• Inspect the part for defects, such as voids or incomplete curing.<br>• Perform mechanical testing to ensure it meets strength and impact resistance requirements.<br><br>By following this approach, you can create a high-quality, lightweight, and durable glass-fiber-reinforced polymer composite panel for a car hood that meets your performance criteria. |

Fig. 9 User Study 1 Prompt 6. The sixth prompt out of six total prompts and corresponding responses provided to users during the first user study. Response 1 was generated by the Composites Guide model and Response 2 was generated by the GPT-4o model.

**Appendix B: Prompts Used in User Study 2**

Prompt 1: What Clamp force do I need for the Injection Molder?

| Response |
|---|
| The BOY 35 EVV injection molding machine operates with a maximum clamping force of 350 kN. This clamping force is designed to secure the mold halves together effectively during the injection process, ensuring proper mold alignment and preventing material leakage. When setting up the machine, ensure that this maximum force is not exceeded to avoid unnecessary strain or damage to the mold and machine components. |

*Fig. 10 User Study 2 Prompt 1.The first prompt out of six total prompts and corresponding response provided to users during the second user study.*

Prompt 2: What are the specifications of the nozzle heater of the Injection Molder according to its manual?

| Response |
|---|
| The BOY 35 EVV machine features a nozzle heater as part of its standard plasticizing unit, which includes five temperature control zones:<br><br>1. **Zone 1, 2, and 3:** Cylinder heating tapes<br>2. **Zone 4:** Plinth heating tape<br>3. **Zone 5:** Nozzle heating tape<br><br>The temperature of these heating tapes is monitored by thermocouples for precise temperature control. For the nozzle heating tape specifically, the maximum operating temperature can reach up to 300°C, and it is essential to ensure proper surface contact with the nozzle body for accurate heat transfer and effective operation. |

*Fig. 11 User Study 2 Prompt 2. The second prompt out of six total prompts and corresponding response provided to users during the second user study.*

Prompt 3: Why is there bubbling in the formed part of the MAAC Thermoformer?

| Response |
| --- |
| Bubbling in the formed parts from the MAAC Thermoformer is typically due to several factors related to material preparation and heating:<br><br>1. **Inconsistent Heating:** If the sheet's core is not heated uniformly, it can lead to excessive stress and potential bubbling. Uniform heating is essential to prevent this issue, ensuring the core and surface reach the same processing temperature.<br>2. **Moisture in the Material:** Certain thermoplastic materials, if not properly dried, can release moisture when heated, creating bubbles or voids in the final product. Ensuring that the plastic sheet or film is adequately dried before forming can help mitigate bubbling.<br>3. **High or Uneven Mold Temperature:** Mold temperature consistency is crucial. If the mold temperature is uneven, some areas may cool too quickly, trapping air or creating localized stresses that result in bubbles.<br><br>To achieve the best results and minimize bubbling, the machine manual suggests using profile heating (even heating across the sheet) and controlling mold temperature carefully. Additionally, rapid vacuum and consistent pressure during forming can reduce internal stresses and potential bubbles in the final part. |

*Fig. 12 The third prompt out of six total prompts and corresponding response provided to users during the second user study.*

Prompt 4: What safety precautions are essential when operating the Injection Molder?

| Response |
| --- |
| When operating the BOY 35 EVV injection molding machine, the following essential safety precautions must be observed:<br><br>1. **Ensure Safety Gate Integrity:** The safety gate is equipped with two interlocking systems, one electrical and one hydraulic. With the safety gate open, the electrical limit switch (-B22) interrupts the control circuit, blocking several valves that control operations such as mold closing, clamping force generation, and injection. This is crucial to prevent accidental operation while the gate is open.<br>2. **De-Energize During Maintenance:** Before performing any maintenance or repair, ensure the entire system, including all electrical, electronic, and hydraulic components, is fully de-energized and depressurized. Only authorized and trained personnel should perform maintenance.<br>3. **Proper Emergency Protocols:** The machine has an emergency stop feature that, when activated, immediately stops operations by closing safety doors and overriding electrical safety circuits. This protocol should be understood and accessible to all operators.<br>4. **Hydraulic Safety:** When the safety gate is open, there should be no hydraulic closing movements. Ensure the presence of DC voltage at specific terminals (K103.13 and X2:L-) to verify the proper function of limit switches affecting safety cartridge valves.<br>5. **Avoid Unauthorized Spare Parts:** Using parts not sourced from the manufacturer may result in system malfunctions and void warranties. These parts can introduce unknown risks and increase diagnostic time during servicing.<br>6. **Limit Handling Around Hydraulic Fluid:** During machine transport or in cases of an accident, be cautious of hydraulic fluid leakage. Proper handling and storage protocols, including securing the machine, should be followed to mitigate such risks.<br><br>Safety Disclaimer<br>Always adhere to the manufacturer's safety protocols and consult the manual for specific safety devices and procedures tailored to the BOY 35 EVV. Unauthorized alterations to the electrical or hydraulic systems may compromise operator safety and machine functionality. |

*Fig. 13 The fourth prompt out of six total prompts and corresponding response provided to users during the second user study.*

Prompt 5: How can I resolve material flow issues during the injection molding process of the Injection Molder?

| Response |
|---|
| To resolve material flow issues during the injection molding process of the BOY 35 EVV, consider the following troubleshooting steps:<br><br>1. **Optimize Injection Pressure:** Ensure that the specific injection pressure is set correctly for the material. For example, polypropylene typically requires around 210 bar. Incorrect pressure settings can lead to insufficient material flow or defects in the molded parts.<br>2. **Check Temperature Settings:** Verify that the barrel, nozzle, and mold temperatures are aligned with the processing requirements of the material. Variations in temperature can result in improper material flow or premature cooling, which affects part quality.<br>3. **Inspect the Screw and Barrel:** If the screw or barrel is worn or damaged, it can hinder material flow. Ensure these components are clean, free of wear, and appropriate for the type of material being processed.<br>4. **Optimize Back Pressure and Screw Speed:** Adjust the back pressure and screw speed to facilitate proper plasticizing and material flow. High back pressure may result in overheating or degradation, while low back pressure can cause inconsistent flow.<br>5. **Avoid Air Traps:** Trapped air in the injection unit can lead to flow issues. Ensure proper venting of the injection unit, especially after long downtimes or transport. The manual suggests at least 10 full stroke cycles for proper venting.<br>6. **Material Preparation:** Some materials require pre-drying or specific storage conditions. Ensure that the material is properly dried and free from contaminants or moisture that may affect its flow properties.<br>7. **Monitor Mold Design and Condition:** Mold geometry, including flow paths and venting, can significantly impact flow. Verify that the mold is correctly designed for the material and that venting is adequate.<br><br>Common Issues<br>- **Inconsistent Flow or Short Shots:** This may indicate incorrect pressure or temperature settings.<br>- **Part Defects or Bubbles:** Often due to trapped air, inadequate mold venting, or improper material temperature.<br><br>Safety Disclaimer<br>Always ensure that any adjustments to pressure, temperature, or mechanical components are conducted by trained personnel following the machine's safety protocols.<br>For further details, refer to the BOY 35 EVV manual's troubleshooting section and operating specifications. |

*Fig. 14 The fifth prompt out of six total prompts and corresponding response provided to users during the second user study.*

Prompt 6: What size sheet is needed for a 12" x 18" mold for the MAAC Thermoformer?

| Response |
| --- |
| For a 12" x 18" mold, the sheet size needed on the MAAC Thermoformer should be slightly larger than the mold dimensions to allow for proper clamping and stretching during the thermoforming process. Generally, a clearance of approximately 1-2 inches on each side is recommended. Thus, a sheet size of about 14" x 20" would be suitable. This ensures that the material can cover the mold completely and accommodate any necessary trimming after forming.<br>For precise details, consult the specific setup guidelines and sheet clamping recommendations within the MAAC Thermoformer manual |

*Fig. 15 The sixth prompt out of six total prompts and corresponding response provided to users during the second user study.*

**Appendix C: Composites Guide and Equipment Assistant Responses**

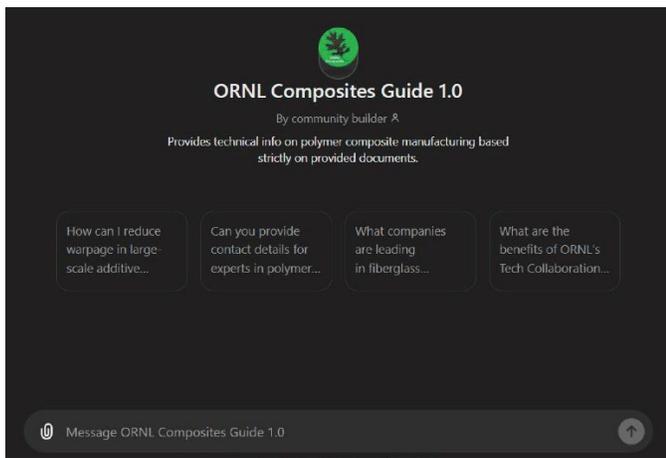 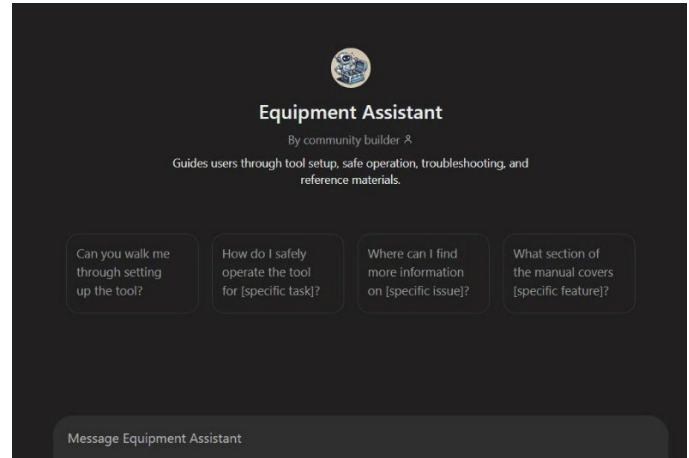

(a) Composites Guide Interface      (b) Equipment Assistant Interface

Fig. 16 The first screen users see when they access the (a) Composites Guide interface and (b) Equipment Assistant interface.

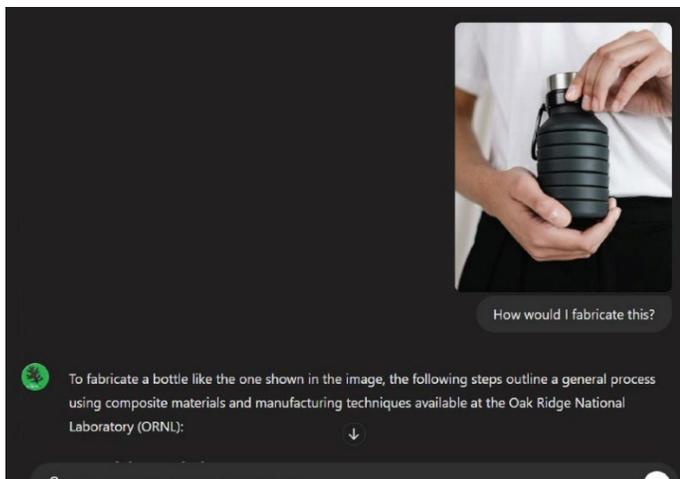 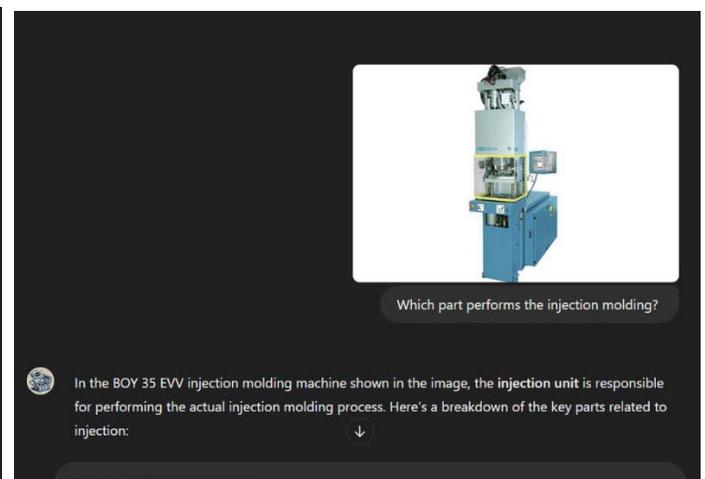

(a) Composites Guide Image Upload      (b) Equipment Assistant Image Upload

Fig. 17 (a) A brief portion of the Composite Guide's response to a user uploading an image of a water bottle and asking "How would I fabricate this?". The model suggests a general procedure based on its knowledge database. (b) A brief portion of the Equipment Assistant's response to a user uploading an image of the Injection Molding Machine and asking "Which part performs the injection molding?". The model provides a breakdown of the key parts of the machine responsible for injection molding.

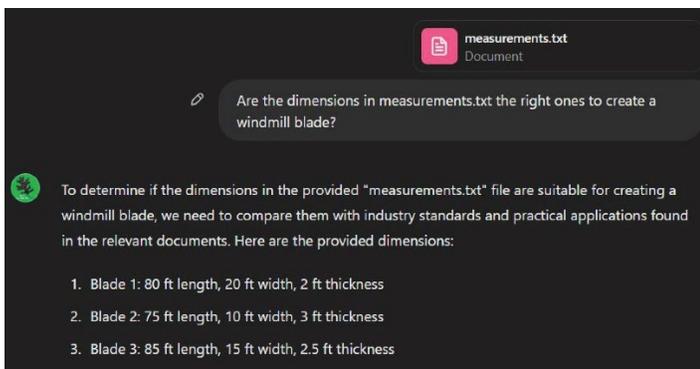 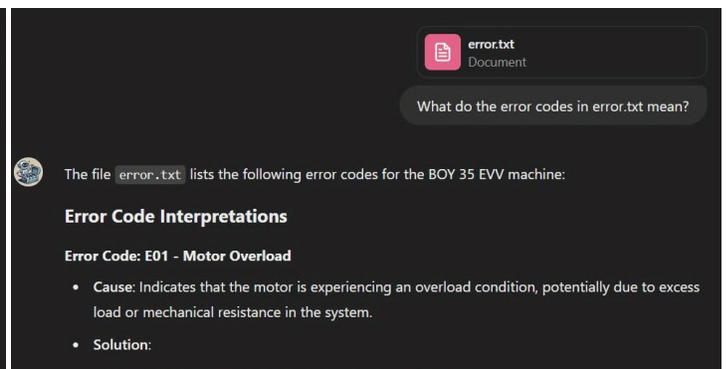

(a) Composites Guide File Upload      (b) Equipment Assistant File Upload

Fig. 18 (a) A brief portion of the model's response to a user uploading a txt file with potential measurements for a windmill blade and asking, "Are the dimensions in measurements.txt the right ones to create a windmill blade?". It correctly accesses the contents of the file and compares it with known standards. (b) A brief portion of the Equipment Assistant's response to a user uploading a txt file with error codes for the Injection Molder and asking, "What do the error codes in error.txt mean?". The model correctly accesses the file and walks the user through an explanation and a potential solution.

# Appendix D: ROUGE Metrics

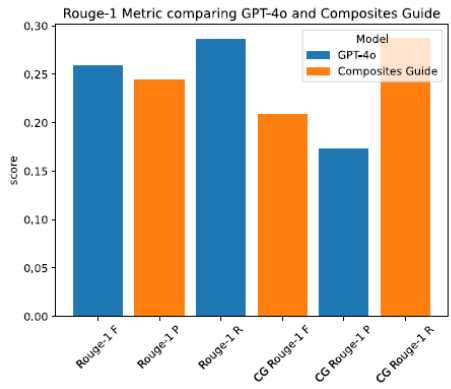

Fig. 19  ROUGE-1 Metric comparing Composites Guide and GPT-4o

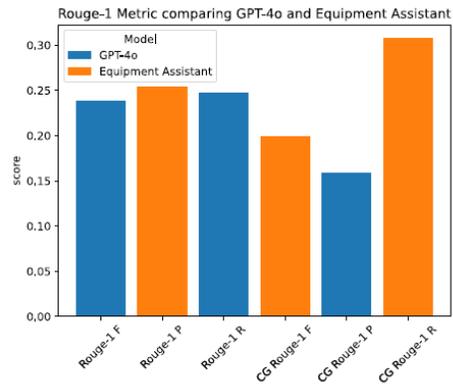

Fig. 22  ROUGE-1 Metric comparing Composites Guide and GPT-4o

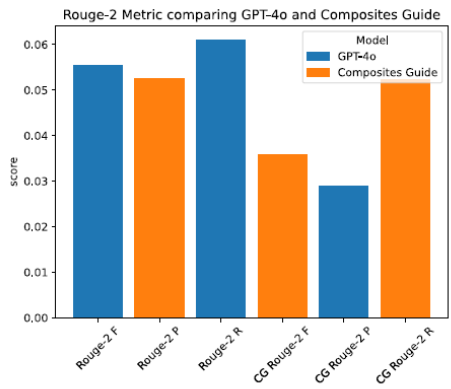

Fig. 20  ROUGE-2 Metric comparing Composites Guide and GPT-4o

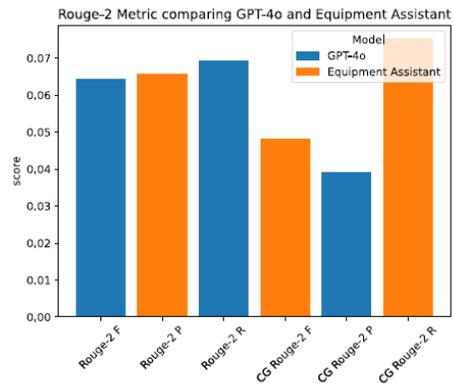

Fig. 23  ROUGE-2 Metric comparing Composites Guide and GPT-4o

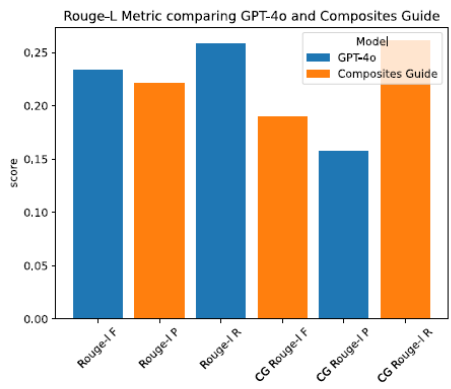

Fig. 21  ROUGE-L Metric comparing Composites Guide and GPT-4o

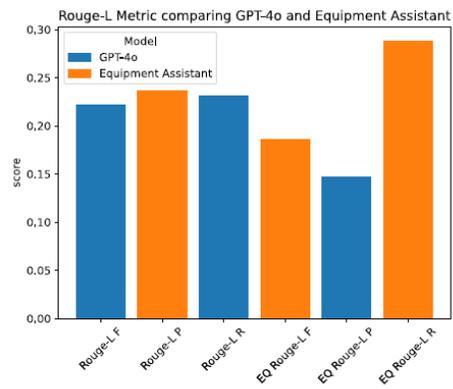

Fig. 24  ROUGE-L Metric comparing Composites Guide and GPT-4o

**Appendix E: BERTScore Metrics**

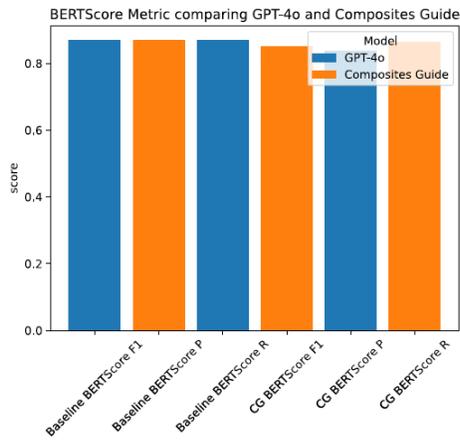

Fig. 25 BERTScore Metric comparing Composites Guide and GPT-4o

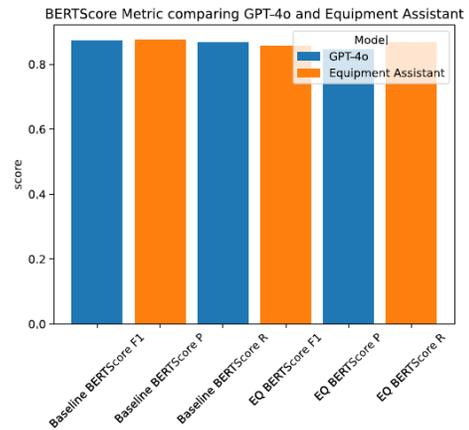

Fig. 26 BERTScore Metric comparing Equipment Assistant and GPT-4o

**Appendix F: Human Rating of GPT-4o and Proposed Models**

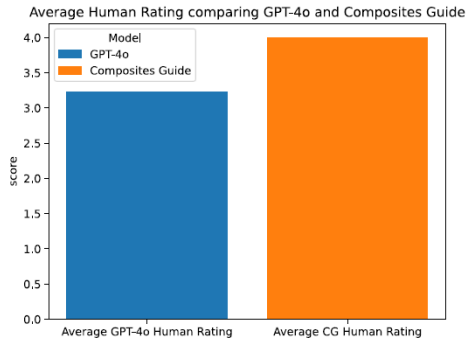

Fig. 27 Human Rating comparing Composites Guide and GPT-4o

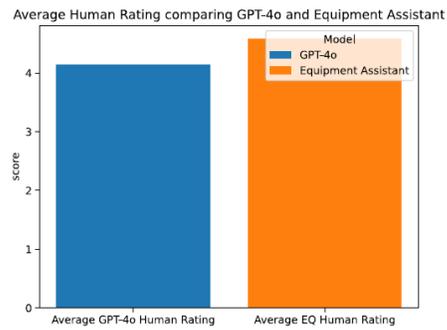

Fig. 28 Human Rating comparing Equipment Assistant and GPT-4o

**Appendix G: First 10 Q&As in the Composites QA Dataset (Total 100 Q&As)**

Question 1: How is the damping ratio affected with excitation amplitude for nonlinear damping systems?

Answer: The experiments showed that the damping ratio increases with an increase in the excitation amplitude. This indicates that at higher excitation levels, more energy is dissipated.

Question 2: How did the cross-sectional shape of the tubes affect the damping ratio?

Answer: The square tube exhibited a higher damping ratio compared to the circular tube. This is attributed to the square tube having more material near the edges, where stresses are higher, which enhances damping. The circular tube showed the lowest damping ratio because most of its material is located near the neutral axis, where bending stresses are minimal.

Question 3: What is the significance of inherent damping in composite materials?

Answer: Inherent damping in composites significantly contributes to the overall damping due to various energy dissipation mechanisms like interfacial slip, micro-cracking, dislocation, and phase boundary effects. Enhancing inherent damping properties can be achieved through proper material selection, loading conditions, and optimizing the cross-section without adding extra weight.

Question 4: What are composites and why are they used?

Answer: Composites are materials made up of two or more materials combined to create superior properties compared to the individual components. They are used because they can achieve desired properties that are not possible with single-phase materials, and their properties can be altered by adding particles to the matrix. These alterations enhance the resulting material's functionality and performance.

Question 5: What is Poisson's ratio, and what is unique about materials with a negative Poisson's ratio (NPR)?

Answer: Poisson's ratio is the ratio of transverse contraction to longitudinal extension in a tensile test. Most traditional materials have a positive Poisson's ratio, meaning they contract when stretched. NPR materials, also known as auxetic materials, expand when stretched and contract when comp

Question 6: What are Triply Periodic Minimal Surface (TPMS) based composites and what are their advantages?

Answer: TPMS-based composites are composites characterized by three-dimensional interconnectivity and periodicity. They divide space into two entwined regions, and the summation of principal curvatures at each point in these structures is zero. TPMS composites possess superior properties compared to counterparts due to the interconnectivity of phases and evade stress concentrations common in truss and strut-based structures, allowing a smooth transfer of load, resulting in better integrity of the composite materials.

Question 7: What is the Level-set approximation and how is it used in generating TPMS-based RVEs?

Answer: The level-set approximation is a methodology used to generate the nodal coordinates that represent TPMS-based RVEs. It is the simplest and most adopted approach for generation, using trigonometric equations where the function represents the surface computed at an isovalue. By plotting the level set to surface $t = 0$, the domain is divided into two subdomains with equal volumes and can be altered by changing the values of t.

Question 8: What are the limitations of existing micromechanical models in predicting the effective properties of composites with negative Poisson's ratio?

Answer: Existing micromechanical models, such as Voigt, Reuss, and Hashin-Shtrikman, do not fully account for the effects of the shape of inclusions (TPMS) and the Poisson's ratio, especially for negative values and higher volume fractions. This leads to deviations in predicted effective properties compared to numerical results17. The mathematical expression given in Eq. (7), along with the constant values provided in Tables 1 and 2, can be useful to estimate the effective properties.

Question 9: What are some applications where superior thermal transport properties are critical?

Answer: Superior thermal transport properties are critical in several engineering applications, including: Electronic packaging design, where efficient thermal management is needed for the effective functionality of devices. Design of protective helmets and body armor, where thermal comfort is essential for users in extreme environments.

Question 10: How do conventional polymeric foams compare to VACNTs in terms of thermal transport properties?

Answer: Conventional polymeric foams generally do not have desirable thermal transport properties due to their random micro and molecular structure. They are also prone to issues like expansion, creep, melting, aging, degradation at high temperatures, and brittle fracture at low temperatures. In contrast, VACNTs exhibit high thermal conductivity, making them more suitable for applications requiring efficient heat transfer.

**Appendix H: First 10 Q&As in the Equipment Manual Assistant QA Dataset (Total 100 Q&As for each manufacturing process – Thermoforming and Injection molding)**

Q1. What is the maximum mold size that the University of Tennessee Machine #4411 MAAC thermoformer can accommodate?

A: The maximum mold size is 72" x 48" (1828mm x 1219mm).

Q2. What is the maximum clamp frame opening size?

A: The maximum clamp frame opening is 73" x 49" (1854mm x 1244mm).

Q3. What are the electrical requirements for operating this thermoforming machine?

A: The machine requires 480V, 60 Hz, with a full load amperage of 180 Amps and a 3-Phase Disconnect size of 250 Amps.

Q4. What type of electrical enclosure does the machine use?

A: The enclosure type is NEMA12 SCIR 5 KA.

Q5. What is the recommended setting for the main pneumatic regulator?

A: The main regulator should be set to 80 Psi / 5.51 Bar / 5.62 k/cm2.

Q6. What is the recommended setting for the eject regulator?

A: The eject regulator should be set to a maximum of 15 Psi / 1.034 Bar / 1.05 k/cm2.

Q7. What is the recommended setting for the IR eye regulator?

A: The IR eye regulator should be set to 5 Psi / .0344 Bar / .035 k/cm2.

Q8. What is the maximum continuous noise level produced by the machine?

A: The maximum continuous noise level is 74 dBA2.

Q9. What is the maximum tool weight that the top or bottom platen can handle?

A: The maximum tool weight for either the top or bottom platen is 1500 lbs.

Q10. What is the assembled weight of the thermoforming equipment?

A: The assembled weight is 12000 lbs.